\documentclass[a4paper, 10pt]{article}
\usepackage[pdfauthor={Tuomas A. Rajala}, hidelinks, colorlinks, citecolor=blue]{hyperref}
\usepackage{authblk}

\usepackage[top=3.3cm, left=2.5cm, right=2.5cm, bottom=3cm]{geometry}
\usepackage{graphicx}
\usepackage[utf8x]{inputenc}
\usepackage{amsmath}
\usepackage{amsfonts}
\usepackage{amssymb}
\usepackage{enumerate}
\usepackage{algorithm}
\usepackage{algpseudocode}
\usepackage{multirow}
\usepackage{tikz}

\usepackage[round]{natbib}

\newcommand{\x}{\mathbf{x}}

\newcommand{\R}{\mathbb{R}}
\newcommand{\E}{\mathbb{E}}
\newcommand{\A}{\mathcal{A}}
\newcommand{\K}{\mathcal{K}}

\newcommand{\diag}{\operatorname{diag}}
\newcommand{\atan}{\operatorname{atan}}
\newcommand{\sign}{\operatorname{sign}}
\newcommand{\glob}{\operatorname{glob}}
\newcommand{\loc}{\operatorname{loc}}

\begin{document}
	
		\title{A review on anisotropy analysis of spatial point patterns}
%
\author[1]{T. Rajala}
\author[2]{C. Redenbach}
\author[3]{A. S\"arkk\"a}
\author[2,4]{M. Sormani}
\renewcommand\Affilfont{\itshape\small}
\affil[1]{Department of Statistical Science, University College London, Gower Street, London WC1E 6BT, UK}
\affil[2]{Mathematics Department, University of Kaiserslautern, 67663 Kaiserslautern, Germany}
\affil[3]{Department of Mathematical Sciences, Chalmers University of Technology and University of Gothenburg, 412 96 Gothenburg, Sweden}
\affil[4]{Fraunhofer Institut f{\"u}r Techno- und Wirtschaftsmathematik, Fraunhofer-Platz 1, 67663 Kaiserslautern, Germany}

\maketitle
		
		\begin{abstract}
			A spatial point pattern is called anisotropic if its spatial structure depends on direction. Several methods for anisotropy analysis have been introduced in the literature. In this paper, we give an overview of nonparametric methods for anisotropy analysis of (stationary) point patterns in $\R^2$ and $\R^3$. We discuss methods based on nearest neighbour and second order summary statistics as well as spectral and wavelet analysis. All techniques are illustrated on both a clustered and a regular example. Finally, we discuss methods for testing for isotropy as well as for estimating preferred directions in a point pattern. 
		\end{abstract}
%
%
%
		


\section{Introduction}

In the early spatial point process literature, observed point patterns were typically small and no replicates were available. Hence, it was natural to assume that the patterns were realizations of stationary and isotropic point processes. In the more recent literature, large and complicated point pattern data with replicates are common and it is not as obvious that stationarity and isotropy hold. Therefore, the validity of these assumptions should be checked prior to the analysis. During the recent years, several authors have paid attention to non-stationarity and currently, it is straightforward to include non-stationarity in many point process models \citep{Myllym2009, Rajala2014530, illian2012, Ang2012, diggle2013, Baddeley2014}. Isotropy of a point pattern, on the other hand, is often still assumed without further checking, and even though several tools have been suggested to detect anisotropy and test for it, they are not so widely used. To make such methods more easily accessible, we have, in this paper, collected methods that can be used to detect anisotropies, test for isotropy, and estimate preferred directions in point patterns. We restrict ourselves to unmarked point patterns and do not discuss orientation of marks.

Directional methods are especially useful for regular point patterns since it can be difficult to visually detect anisotropy in such patterns. One example is the amacrine cells data, see Figure \ref{fig:examplesClustered2D} (right), which consist of locations of 'on' cells and 'off' cells. These data have been analyzed by several authors assuming stationarity and isotropy. However, it was recently detected by \citet{Wong2016} that both the marginal 'on' and 'off' patterns as well as the unlabeled pattern show some signs of anisotropy. Locations of air bubbles in polar ice are another example of regular anisotropic patterns \citep{Redenbach2009,Raj16}. Deep down in an ice sheet, the ice, and therefore the air bubble pattern, are deformed. By using directional analysis, we can learn more about the deformation and provide useful information to the glaciologists. Detecting anisotropies visually in the ice samples is especially hard since the air bubble patterns are not only regular but also in 3D. In clustered patterns, the shape and direction of clusters can reveal anisotropies but directional analysis is needed, for example, to estimate the preferred direction of the clusters. In a series of papers on spectral analysis \citep{Renshaw1983, Ford1984,Renshaw1984}, the authors emphasize that in ecological data, especially in growth processes, directional components are common and assuming isotropy is not acceptable.
 
Anisotropy can be caused by several mechanisms. In the existing literature, directional analysis has mainly focused on two types of anisotropy: geometric anisotropy, where anisotropy is caused by a linear transformation of a stationary and isotropic process, and increased intensity of points along directed lines. Geometric anisotropy has been considered both for clustered point patterns, such as the Welsh chapel data \citep{Mugglestone1996b, Moller2014}, and for regular point patterns, such as the earlier mentioned amacrine cells and air bubble data sets. The {\it Ambrosia dumosa} dataset is an example of an anisotropic point pattern with increased intensity along directed lines \citep{Rosenberg2004}. In this paper, we restrict ourselves to these two types. 

We concentrate on reviewing methods that are generally applicable for any point pattern and do not require a specific model assumption. Therefore, we focus on methods that are based on spatial summary statistics, such as the nearest neighbor distance distribution function, Ripley’s $K$ function and the pair correlation function, spectral analysis, and wavelet analysis. Several tests for isotropy have been proposed based on these methods. Some of these tests are asymptotic, some based on Monte Carlo simulations and some rely on replicated data. Directional analysis based on summary statistics has been used both in 2D and 3D, whereas analysis based on spectral analysis and wavelets has so far been introduced only in 2D. Additionally, wavelet analysis discussed in the literature concentrates on situations where we have increased intensity along directed lines. We mention some models for anisotropic point patterns but will not discuss them thoroughly as they are typically tailor made for a specific situation and data, with particular models for the point location processes.

To illustrate how a typical result of each method looks like, we apply all presented methods to two  simulated 2D data sets, a regular compressed point pattern and a clustered pattern with increased intensity along directed lines. Both point patterns are realizations of stationary point processes. The regular pattern is an example of geometric anisotropy and is comparable to  the amacrine cells and the air bubble data. The clustered pattern has similar features as the {\it Ambrosia dumosa} data and the pyramidal cells data sets considered in \citet{Rafati2016}. The Welsh chapels data set, an example of a geometrically anisotropic clustered pattern, is not covered by these two examples. Based on the analysis of the simulated example point patterns, we make some observations about how suitable the different methods seem to be in these particular cases. However, a much more thorough analysis of the methods would be needed to be able to give any general recommendations.  

Introduction to our notation is given in Section \ref{sec:Defs}. In Section \ref{sec:Mechanisms},  we describe the two mechanisms causing anisotropy included in this paper, geometric anisotropy and clustering along directed lines. Methods based on nearest neighbour and second order summary statistics are recalled in Section \ref{sec:SummaryStatistics}, spectral analysis in Section \ref{sec:SpectralAnalysis}, and wavelets in Section \ref{sec:Wavelets}. 
Finally, we give an overview of the tests for isotropy presented in the literature in Section \ref{sec:Testing} and conclude by discussing our findings and future work.

\section{Definitions and notation}
\label{sec:Defs}
In this section, we give some basic definitions concerning point processes. Let $\x=\{x_1,...,x_n\}$ be a point pattern observed in a window $W\subseteq \R^d$ with $n>2$ points. We assume that $\x$ is obtained by intersecting a realization of a simple (no multiple points) point process $X$ with $W$, and that $W$ is bounded with volume $|W|$. 

The point process $X$ is stationary, if its distribution is invariant under shifts in $\R^d$. If its distribution is invariant under rotations around the origin, the process is isotropic. Unless stated otherwise, we will assume that the point process $X$ is stationary. The task is then to detect and quantify anisotropies in the observed point pattern. Note that some of the methods described below can be generalized to second order intensity reweighted stationary point processes \citep{Baddeley2000}, see Section \ref{sec:SecondOrder_Nonstationary}.

The point process $X$ can be regarded as a random collection of points (as above) or as a counting measure on $\R^d$, also denoted by $X$. In the latter case, $X(B)$ is the random number of points of $X$ in a Borel set $B$ in $\R^d$. In particular, $X(W)=n$. The intensity function of $X$ is defined as
$$
\lambda(x)= \lim_{|dx| \to 0} \frac{\E[X(dx)]}{|dx|}.
$$
In the stationary case, $\lambda(x)\equiv\lambda$, where $\lambda > 0$ is a constant. The Palm distribution of $X$ will be denoted by $P_x$ for any $x\in \R^d$. Heuristically, it can be interpreted as the conditional distribution of $X$ given $x\in X$. The corresponding expectation is denoted by $\E_x$. For stationary $X$, it suffices to consider only $P_o$ and $\E_o$ where $o\in \R^d$ is the origin.

In the directional analysis, it is often convenient to use polar coordinates in 2D and spherical coordinates in 3D. The 2D Cartesian coordinates $(x,y)$ can be represented in polar coordinates as 
$$
r=\sqrt{x^2+y^2} \in [0, \infty) \quad\text{and}\quad \phi=\arctan(y/x) \in[0, 2\pi)
$$ 
and the 3D coordinates $(x,y,z)$ as 
\begin{align*}
r&=\sqrt{x^2+y^2+z^2} \in [0, \infty),\\
\phi&=\arctan(y/x) \in [0, 2\pi), \quad\text{and}\\ \theta&=\cos^{-1}\left(\frac{z}{\sqrt{x^2+y^2+z^2}}\right)=\cos^{-1}(z/r)\in [0, \pi].
\end{align*}
We use the convention 
that  the  2D angle goes anti-clockwise from the $x$-axis. Formally, we set $\arctan(y/x)=[(\atan(y/x)+2\pi)\text{ mod } 2\pi]$ where $\atan(y/x)$ is $\tan^{-1}(y/x)$ if $x>0$, $\tan^{-1}(y/x)+\sign(y)\pi$ if $x<0$, $\sign(y)\pi/2$ if $x=0$, and left undefined if $x=y=0$. 

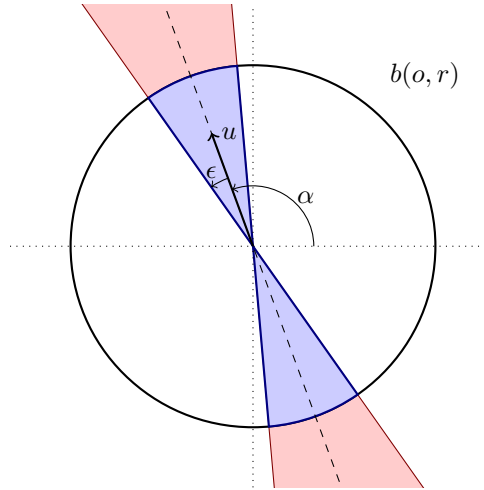
\begin{figure}
\begin{center}
        
		
		

\begin{tikzpicture}[scale=0.8]
		\clip (-4,-4) rectangle (4,4);
		
		\filldraw[fill=red!20!white, draw=red!50!black, rotate=95] (0,0) -- (7cm,0) arc(0:30:7cm) -- cycle;
		\filldraw[fill=red!20!white, draw=red!50!black, rotate=-85] (0,0) -- (7cm,0) arc(0:30:7cm) -- cycle;
		\draw[thick] (0,0) circle (3cm);
		\filldraw[fill=blue!20!white, draw=blue!50!black, rotate=95, thick] (0,0) -- (3cm,0) arc(0:30:3cm) -- cycle;
		\filldraw[fill=blue!20!white, draw=blue!50!black, rotate=-85, thick] (0,0) -- (3cm,0) arc(0:30:3cm) -- cycle;
	
        \draw[->, thick, rotate = 110](0,0) -- (2cm,0) node [right]{$u$};

		\draw[->] (1cm,0) arc (0:110:1cm) node[midway, right]{$\alpha$};	
        \draw[->,rotate=110] (1.2cm, 0) arc (0:15:1.2cm) node[above]{$\epsilon$};
		\draw[dashed, rotate=110] (-6,0) -- (6,0);
		
		\draw (45: 4cm) node {$b(o,r)$};
		\draw[dotted] (-6,0) -- (6,0);
		\draw[dotted] (0,-6) -- (0,6);
\end{tikzpicture}

\caption{The sets $C(\alpha, \epsilon)$ (red and blue) and $S(\alpha, \epsilon, r)$ (blue) when $\alpha=\alpha(u)=3\pi/5, \epsilon=\pi/12$, and $r = 1.5$. 
} 
\label{fig:DirectionalSets}
\end{center}
\end{figure}

Let $b(x,r)$ denote a ball in $\R^d$ with center $x$ and radius $r>0$. For a unit vector $u \in \R^d$, i.e. an element of the unit sphere $S^{d-1}=\{x\in\R^d:||x||=1\}$, the polar/spherical coordinates are $(r,\alpha)=(1,\alpha)$, where $\alpha=\alpha(u)$ gives the angle $\phi$ in 2D and angles $\phi$ and $\theta$ in 3D. The infinite double cone with a central axis spanned by $u$ and with the opening half angle $\epsilon>0$ is denoted by $C(u, \epsilon)$. Furthermore, we set $S(u, \epsilon, r)= C(u, \epsilon) \cap b(o,r)$. Equivalently, when considering angles we will write $C(\alpha, \epsilon)$ and $S(\alpha, \epsilon, r)$ with $\alpha=\alpha(u)$. An illustration is shown in Figure \ref{fig:DirectionalSets}.

We denote the Minkowski sum of two sets $A$ and $B$ in $\R^d$ by
$$
A \oplus B= \{ a+b : a \in A, b \in B\}.
$$ 
Then, $W_x= W \oplus \{x\}$ is the translation of the window $W$ by a point $x \in \R^d$. The Minkowski difference of $A$ and $B$ is defined as
$$
A \ominus B = \{ x \in \R^d  : B_x \subset A\}.
$$

We will write $1(\mathcal{E})$ for the indicator function taking the value 1 if event $\mathcal{E}$ is true and 0 otherwise.

\section{Anisotropy mechanisms}
\label{sec:Mechanisms}

A point pattern can be anisotropic in different ways. Here, we recall two types of anisotropy, geometric anisotropy and anisotropy caused by oriented clusters, that are most common in the point process literature.

\subsection{Geometric anisotropy}
\label{sub:GeometricAnisotropy}
Let $X_0$ be a stationary and isotropic point process. Define a point process $X$ by the transformation $X=TX_0= \{Tx:x \in X_0\},$  where $T:\R^d \to \R^d$ is an invertible linear mapping. The corresponding $d \times d$-matrix will also be denoted by $T$. Since $X_0$ is isotropic, we can decompose the mapping into two matrices, $T=RC$, where $R$ is a rotation matrix and $C$ is a diagonal scaling matrix that compresses and stretches the dimensions. Note that the definition excludes shear. 

If $C$ is not a multiple of the identity matrix, the resulting process $X$ can be anisotropic. Furthermore, it has the following properties:
\begin{enumerate}
\item
$X$ is stationary.
\item
The connection between the counting measures is $X(B)=X_0(T^{-1}B)$ for any Borel set $B \subset \R^d$.
\item
The intensities of $X$ and $X_0$ are related via $\lambda_X= \det(T^{-1})\lambda_{X_0}$.
\item
If $X_0$ is a stationary Poisson process with intensity $\lambda>0$, then $X$ is a stationary Poisson process with intensity $\det(T^{-1})\lambda$. In particular, $X$ is also isotropic.
\end{enumerate}

In \cite{Moller2014}, the type of anisotropy defined above is called geometric anisotropy. The term is borrowed from geo-statistics: The distance $r=x-x'$ of two points is transformed into $r_T=(x-x')^T T^TT(x-x')$, so that, in analogy to the covariance function in geo-statistics, the second order properties of $X$ depending on $r=x-x'$ can be expressed as the second order properties of $X_0$ depending on $r_{T^{-1}}$.
An alternative name for the transformation is elliptical anisotropy: For any sphere $S=\{x:x^Tx=m\}$ in $\R^d$ we have that $TS=\{x:x^T(TT^T)^{−1}x=m\}$ is an ellipsoid (ellipse in 2D). Note that \cite{Moller2014} formulate the transformation in terms of $\Sigma=TT^T$.

\begin{figure}[h!]
	\begin{center}
		\includegraphics[width=0.3 \textwidth]{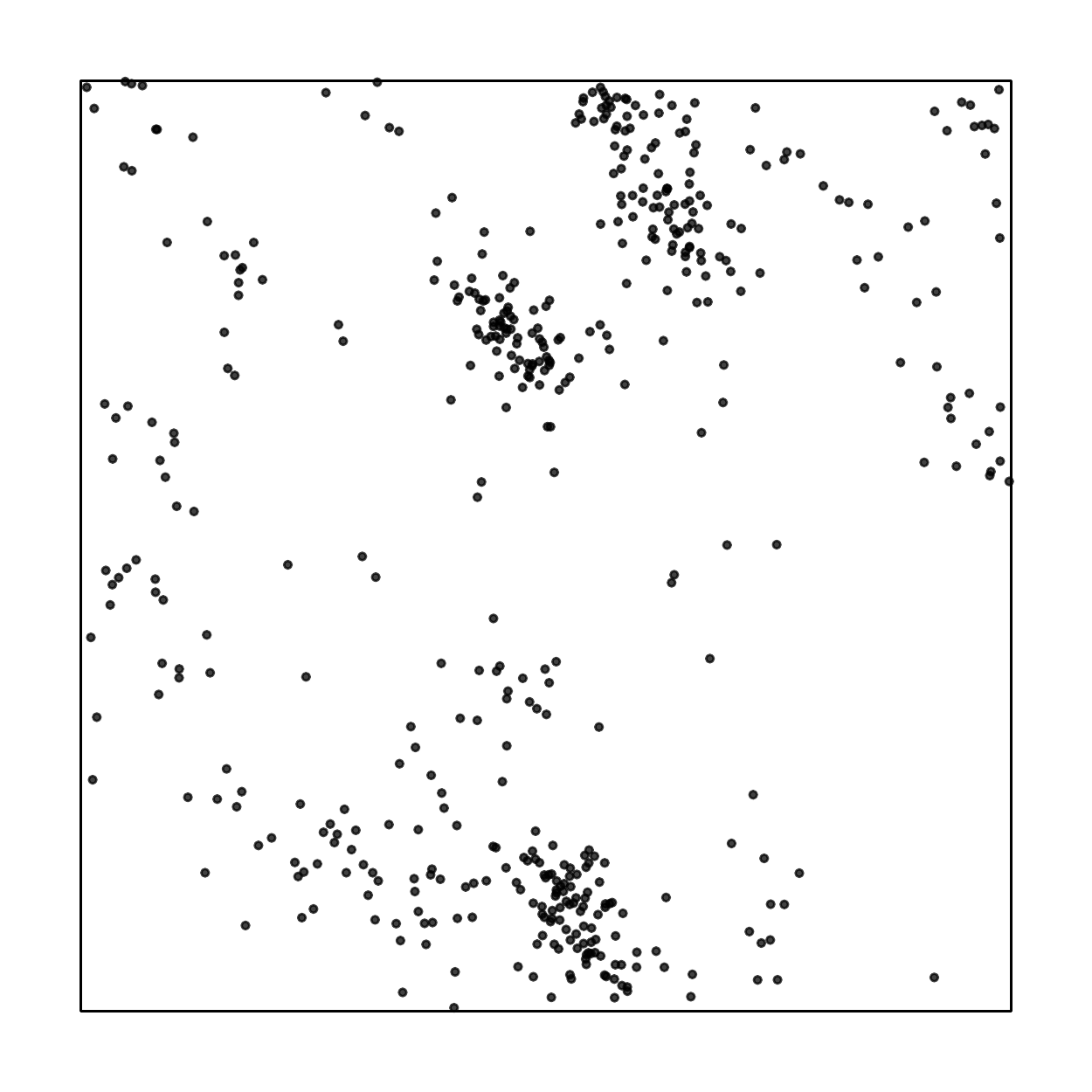}
	 	\includegraphics[width=0.3 \textwidth]{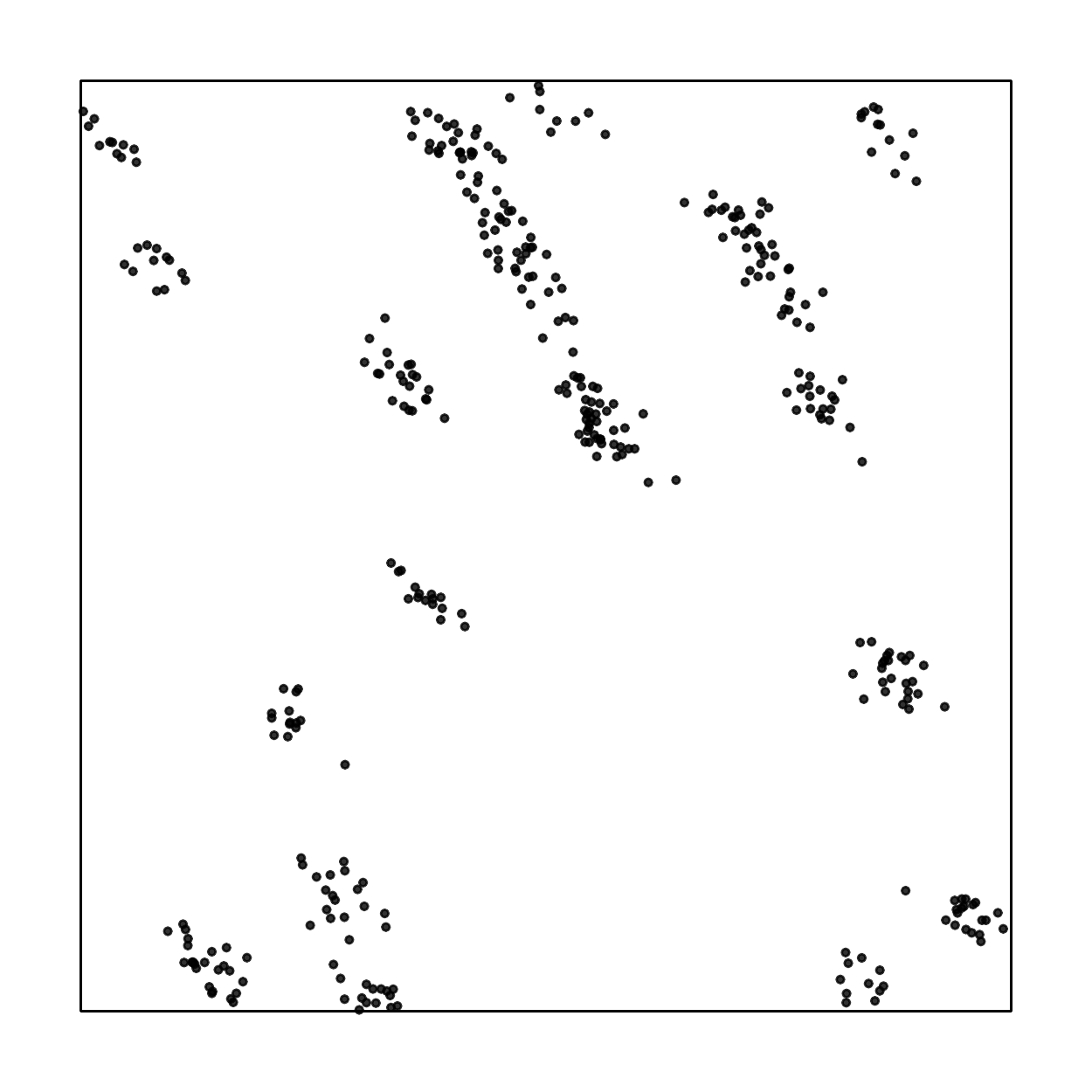}
		\includegraphics[width=0.3 \textwidth]{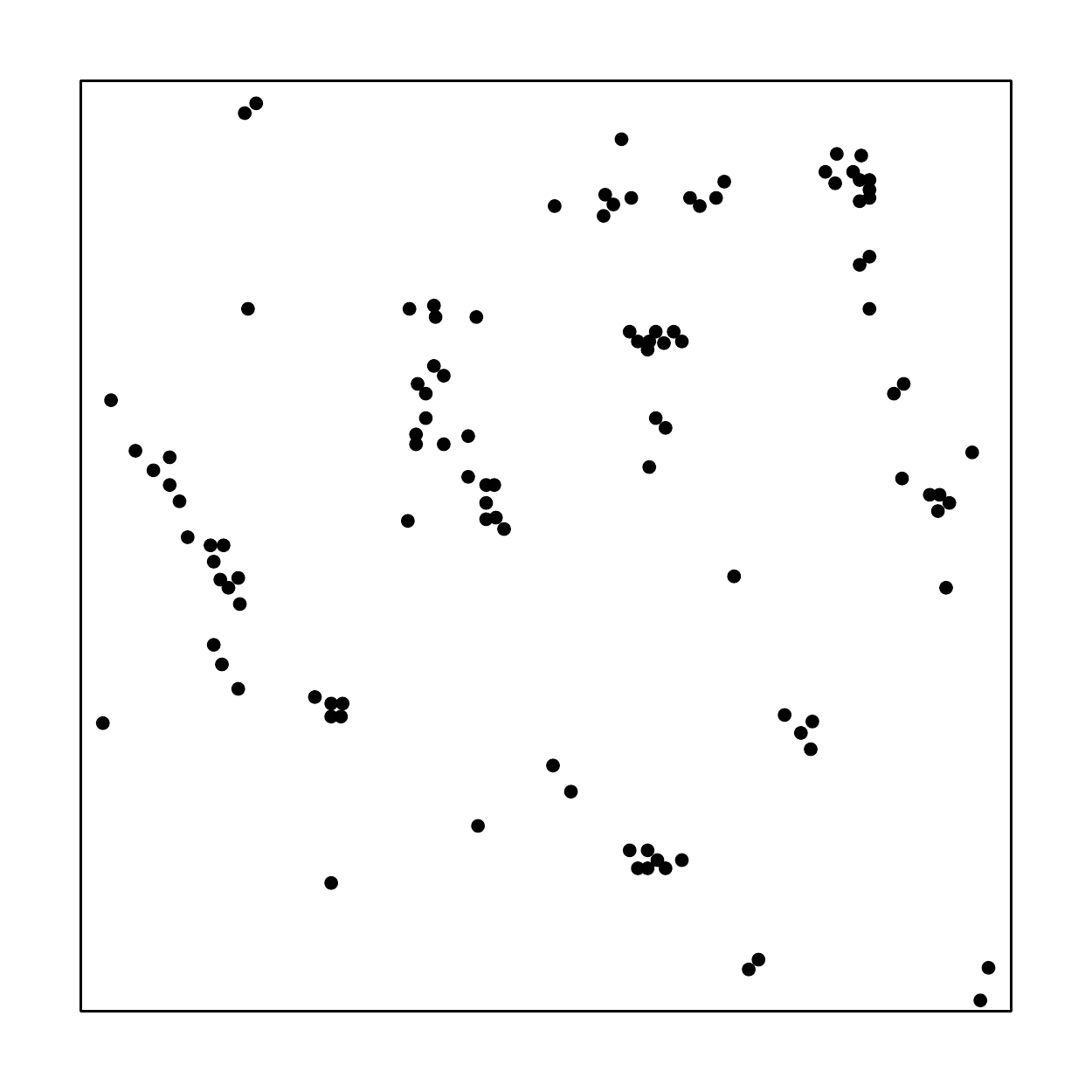}
	\end{center}
	\caption{Linearly transformed realisations of a log-Gaussian Cox process (left) and a Thomas process (middle), and the Welsh chapels data (right). }
	\label{fig:ExamplesGeometricAnisotropyClustered}
\end{figure}

Geometric anisotropy has been studied as a model of anisotropy for both clustered and regular point patterns. Cluster processes were considered in \cite{Moller2014},  \cite{Guan2006}, and \cite{Wong2016}. \cite{Moller2014} consider log-Gaussian Cox processes and shot noise Cox processes, \cite{Guan2006} and \cite{Wong2016} use Poisson cluster processes with elliptic clusters generated by an anisotropic multivariate normal distribution, i.e. anisotropic Thomas processes. Note that, due to Property 4 above, linear transformations of Poisson cluster processes are Poisson cluster processes with transformed clusters.
An example of a real data set with this structure is the Welsh chapel data discussed in \cite{Moller2014} and \cite{Mugglestone1996}. This data set as well as some model realisations are shown in Figure \ref{fig:ExamplesGeometricAnisotropyClustered}. 

\begin{figure}[h!]
	\begin{center}
	\includegraphics[width=0.3\textwidth]{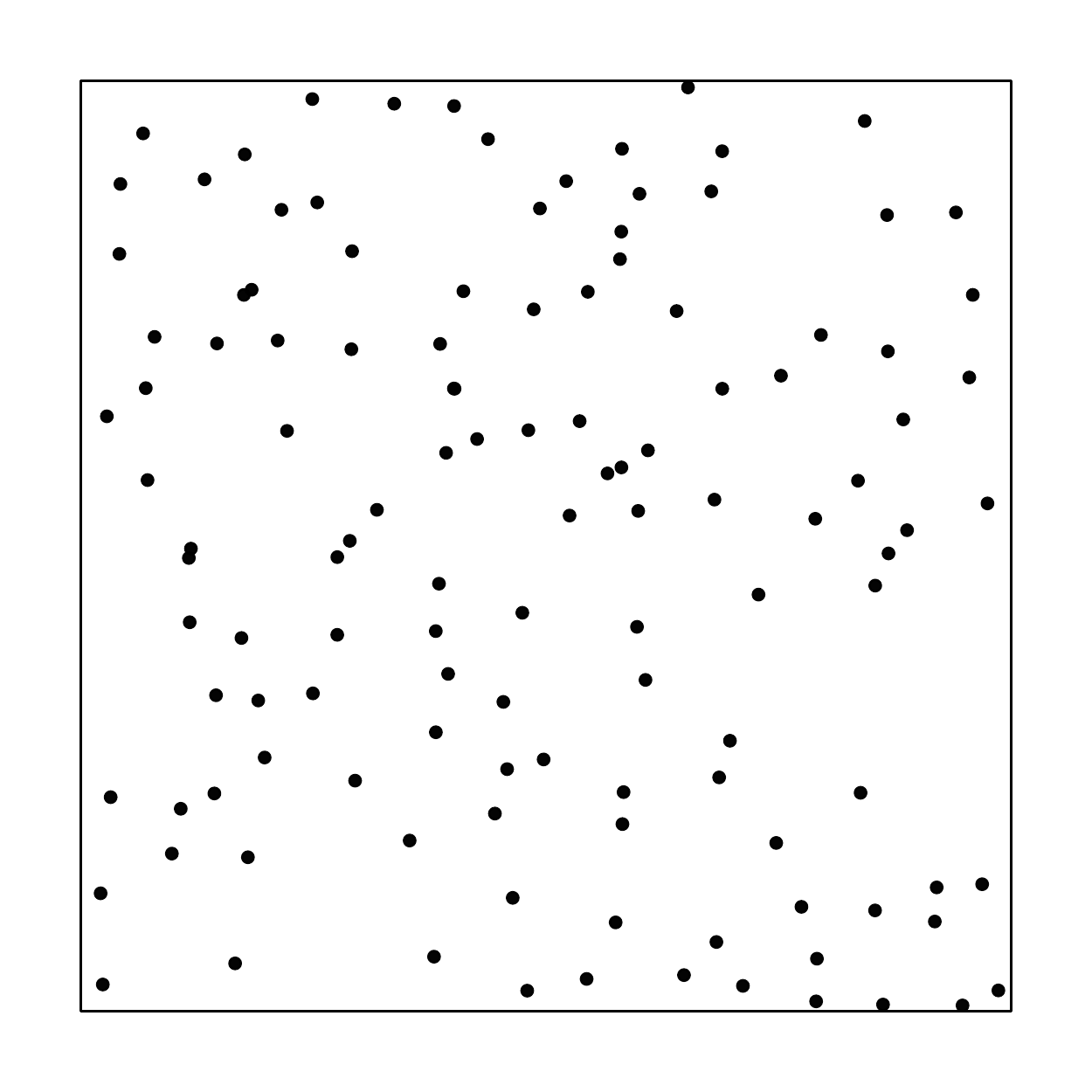}
	\includegraphics[width=0.3\textwidth]{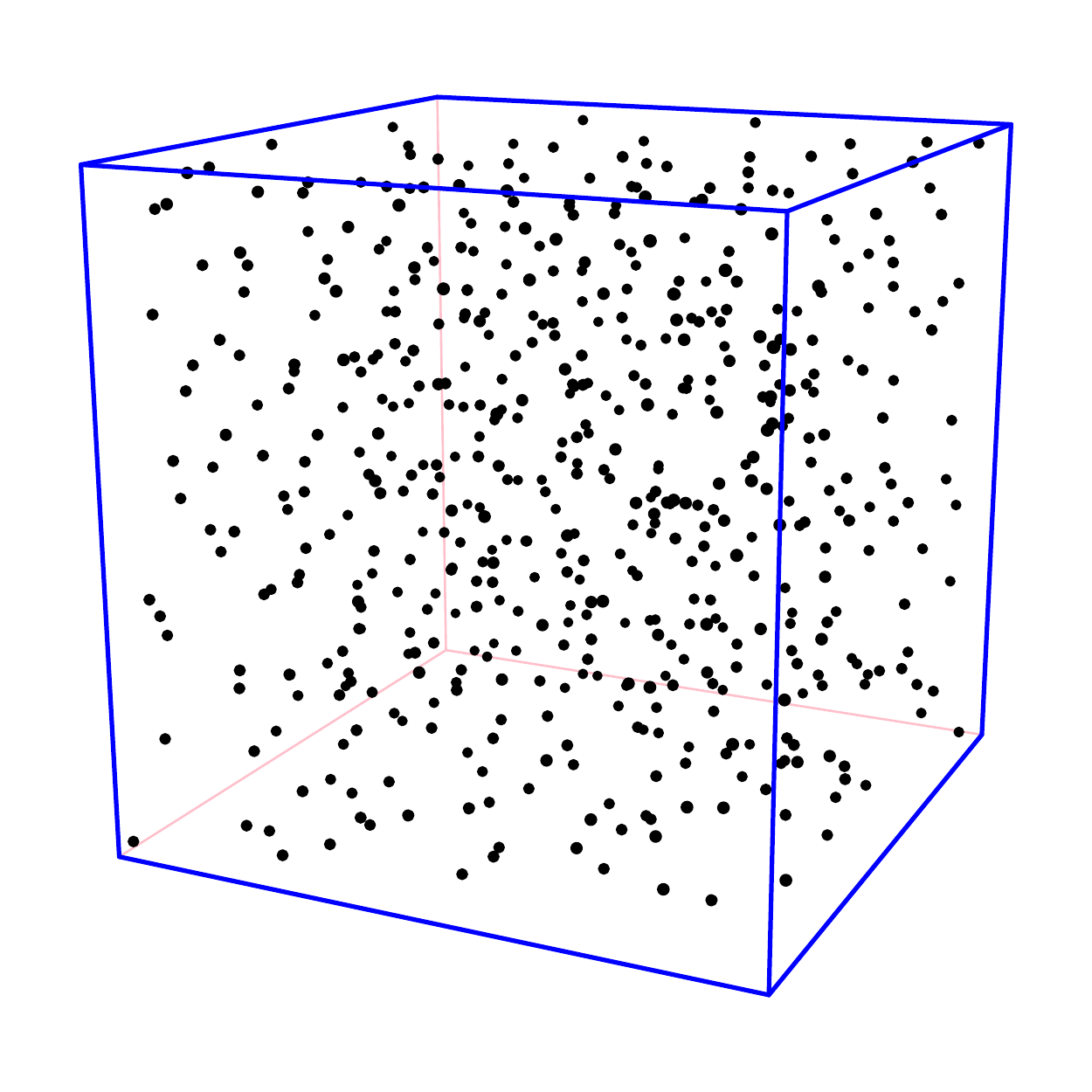}
	\includegraphics[width=0.35\textwidth]{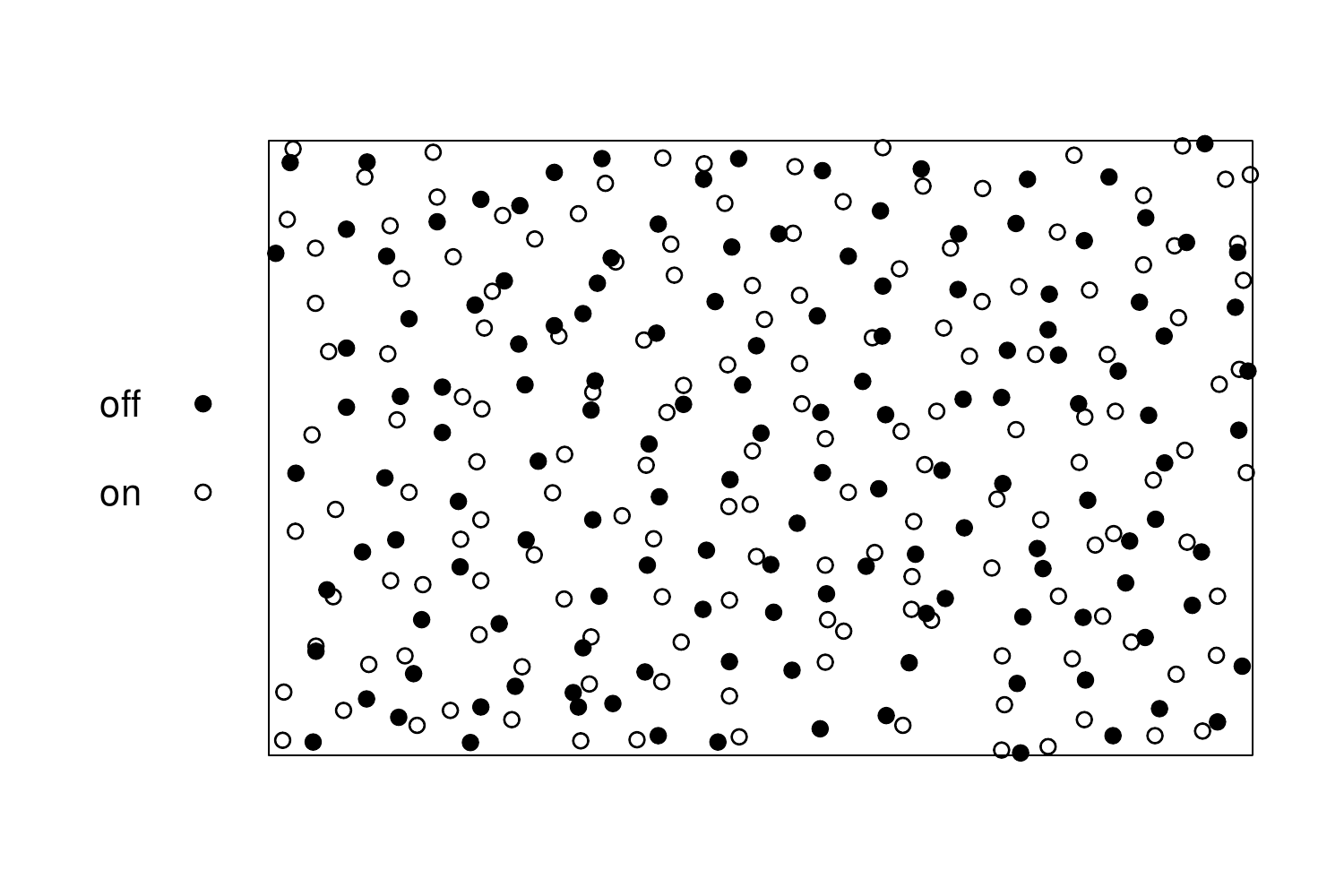}
	\end{center}
	\caption{A realization of a compressed regular point process (left), locations of air bubbles in polar ice (middle), and locations of 152 amacrine cells labelled 'on' and 142 cells labelled 'off' (right). }
	\label{fig:ExamplesGeometricAnisotropyRegular}
\end{figure}

Linear transformations of regular point patterns are studied in \cite{Redenbach2009}, \cite{Raj16} and in \cite{Wong2016} (see Figure \ref{fig:ExamplesGeometricAnisotropyRegular}, left). In the simulation study in \cite{Wong2016}, the regular case is represented by anisotropic Gibbs hard-core processes, in \cite{Raj16} by transformed Strauss processes, and in \cite{Redenbach2009} by transformed Matern hard core processes. As an example of real data,  \cite{Redenbach2009} and \cite{Raj16} study the locations of air bubbles in polar ice (see Figure \ref{fig:ExamplesGeometricAnisotropyRegular}, middle), while in \cite{Wong2016}, amacrine cells in the retina of a rabbit are investigated (\cite{Diggle1986}, see Figure \ref{fig:ExamplesGeometricAnisotropyRegular}, right).

\subsection{Oriented clusters}

\begin{figure}[h!]
	\begin{center}
		\includegraphics[width=0.3 \textwidth]{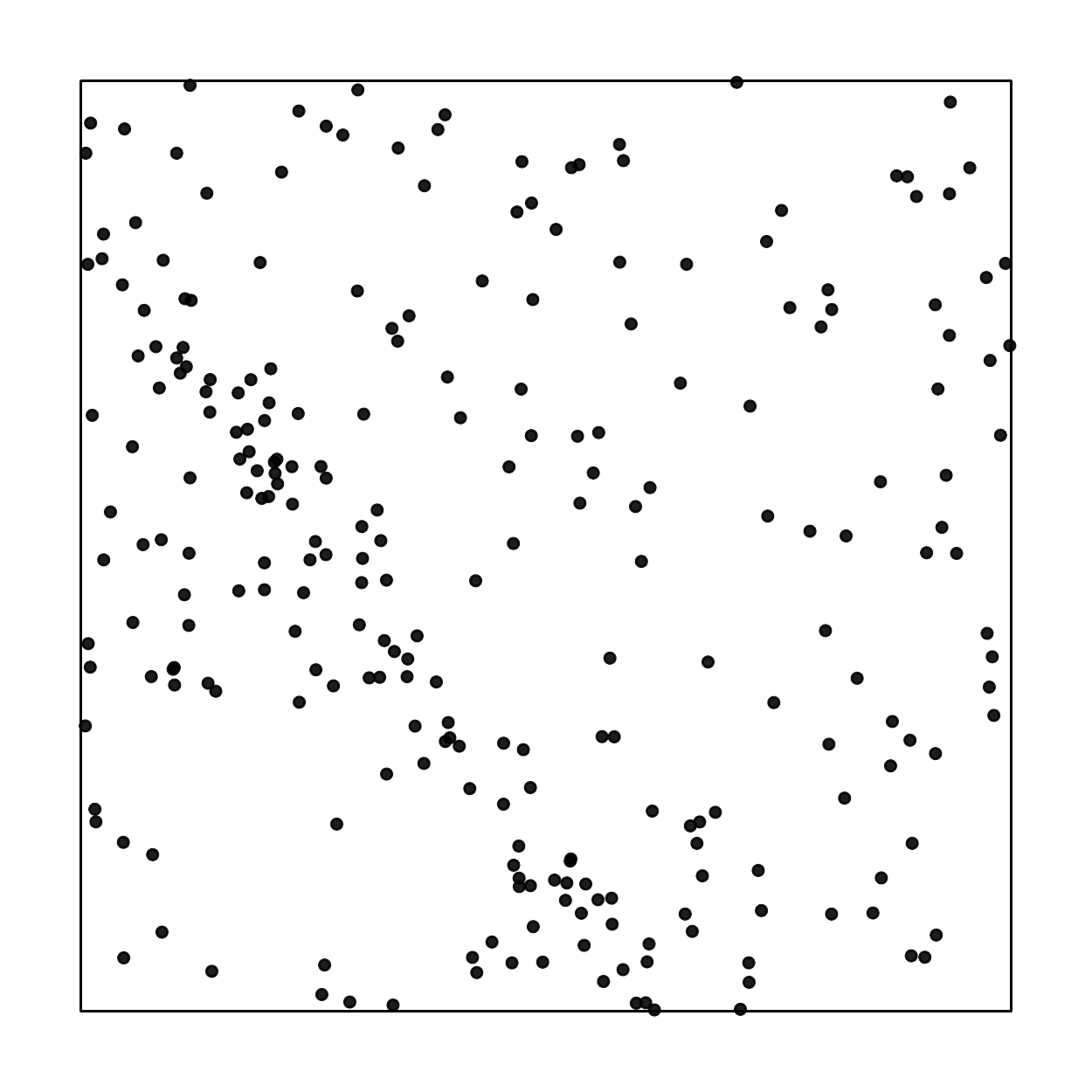}
		\includegraphics[width=0.3 \textwidth]{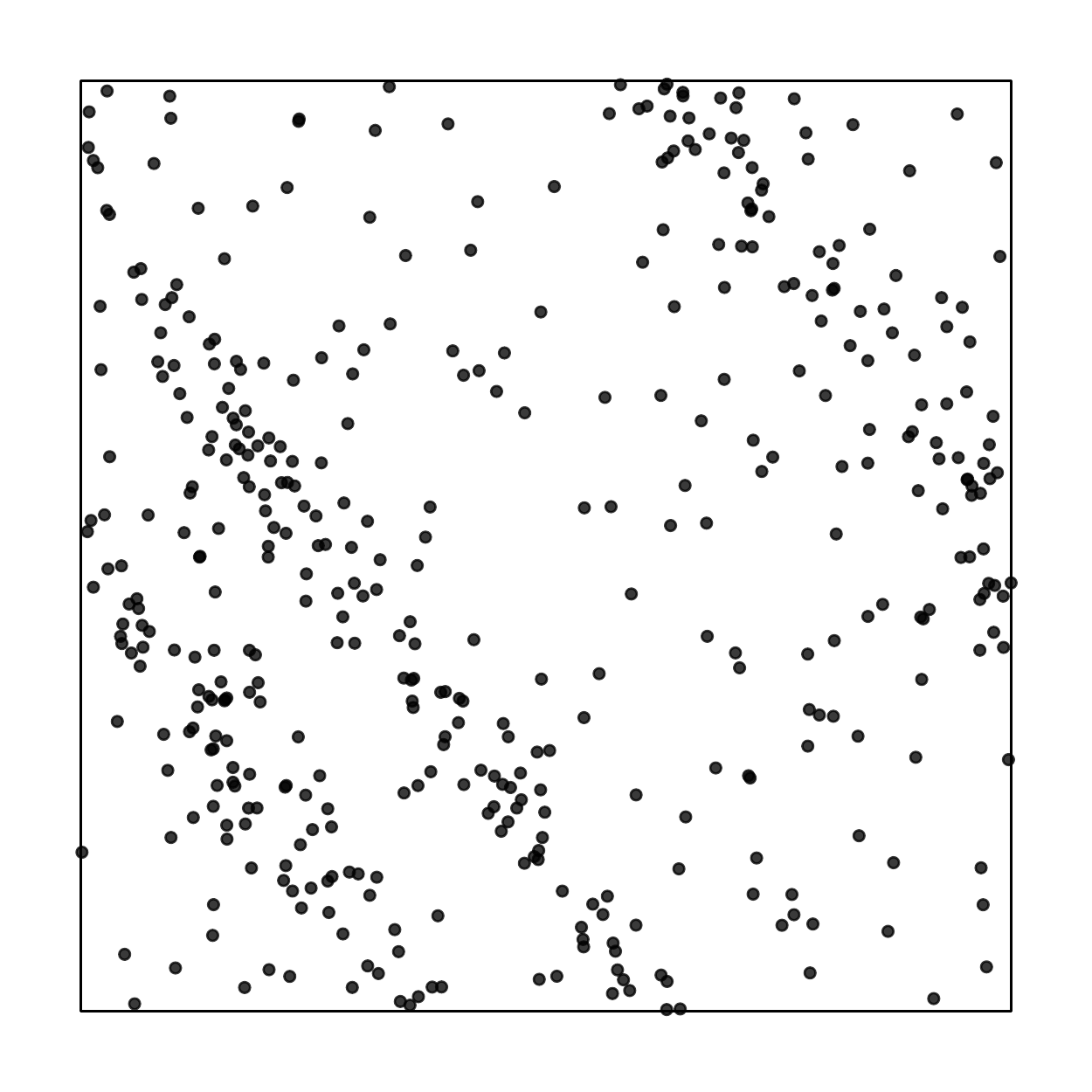}
		\includegraphics[width=0.3 \textwidth]{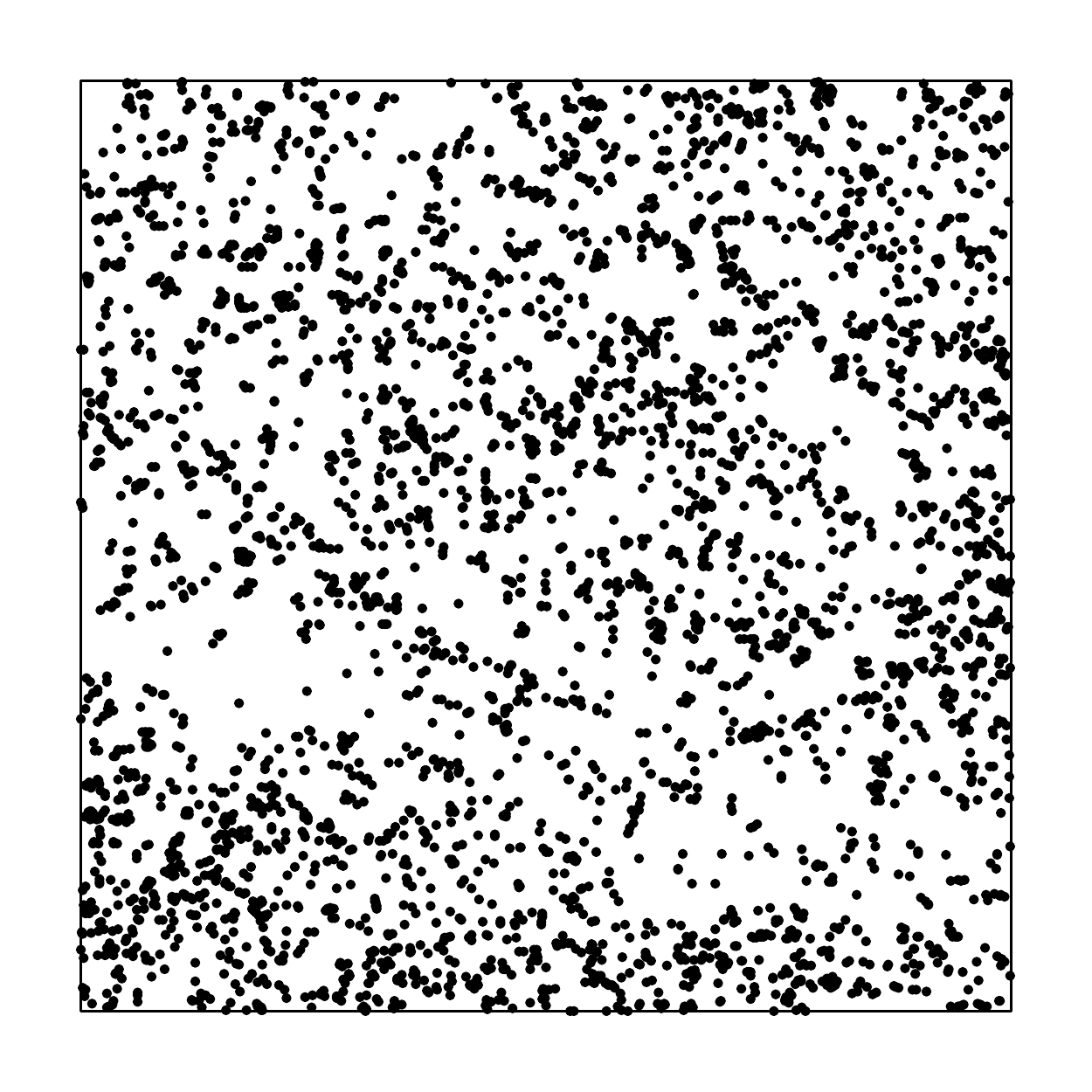}
\end{center}
	\caption{A superposition of a stationary Poisson process with a Poisson process whose intensity is concentrated around a line (left), and a version of this process with three lines which can be considered stationary (middle). Ambrosia dumosa data (right).}
	\label{fig:examplesClustered2D}
\end{figure}

Cluster processes with bounded anisotropic clusters were already considered in the last section. Further typical examples found in the anisotropy literature are Poisson processes with increased intensity along directed lines, see for instance \cite{Rosenberg2004}. These processes can be considered stationary if the distribution of line locations is stationary, e.g. given by a stationary Poisson line process (see Figure \ref{fig:examplesClustered2D}, left, middle). 
A real data set showing such structure is the {\it Ambrosia dumosa} data \citep{Miriti1998}, shown in Figure \ref{fig:examplesClustered2D} (right). To model point patterns with clustering around oriented line segments, models introduced in \cite{Lawson2007} can be useful.

\begin{figure}[h!]
	\begin{center}
	\includegraphics[width=0.3 \textwidth]{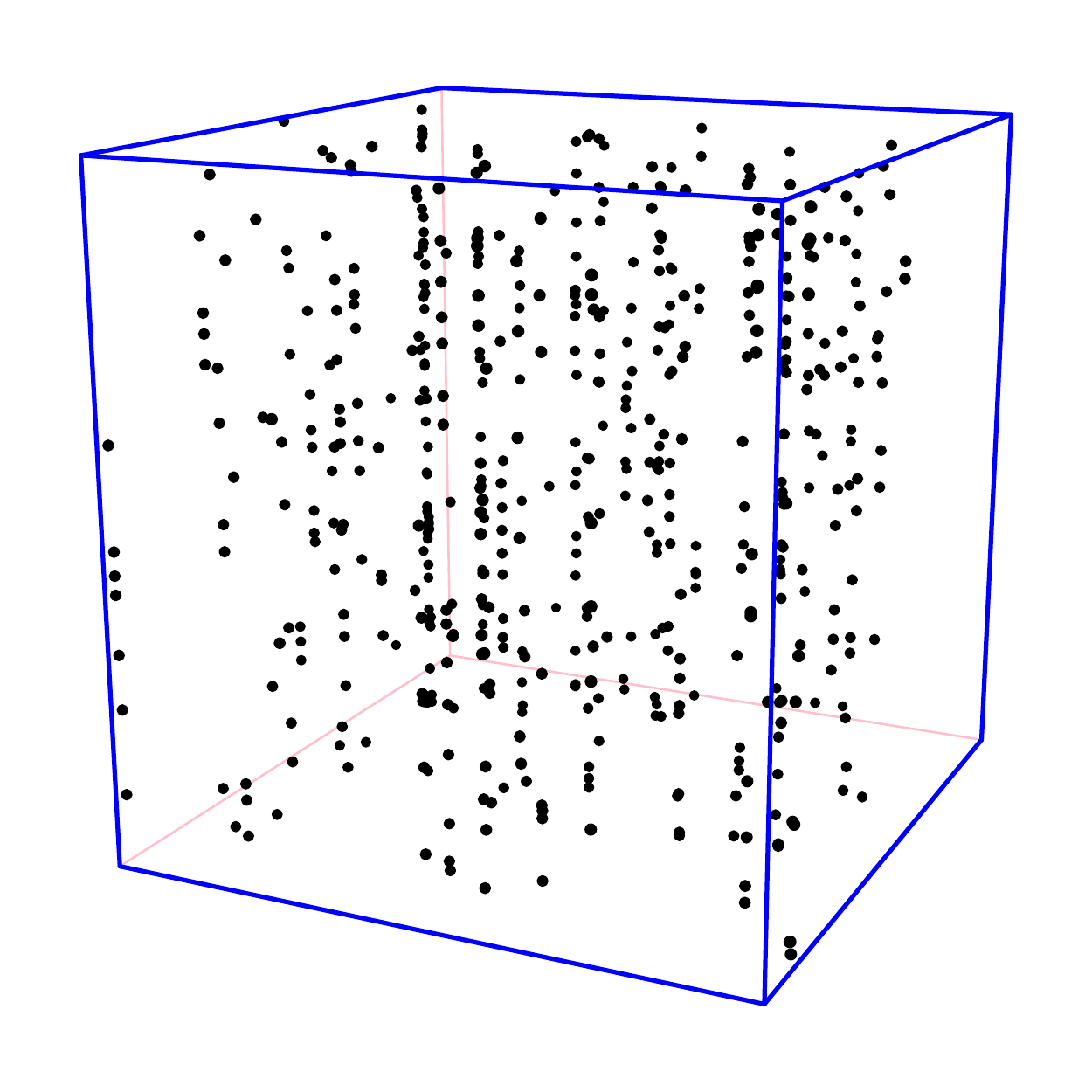}
	\includegraphics[width=0.5 \textwidth]{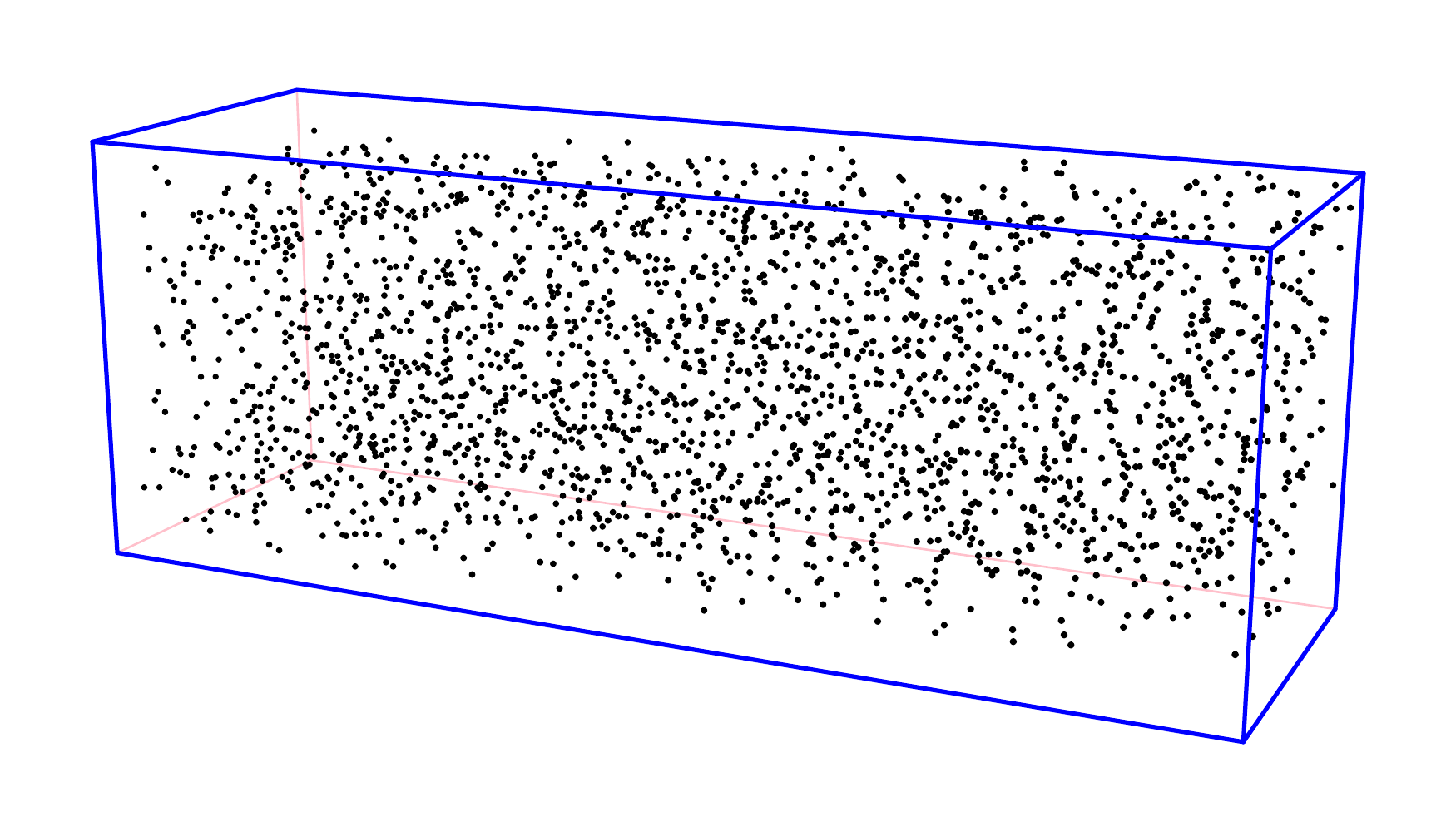}
	\end{center}
	\caption{A realisation of a Poisson line cluster point process (left), and locations of nucleoli of pyramidal cells (right). 
    }
	\label{fig:examplesClustered3D}
\end{figure}

An example in 3D can be found in \cite{Rafati2016} where locations of pyramidal cells in the brain are investigated. The minicolumn hypothesis in neuroscience states that these cells are organized in parallel columns which results in an anisotropic arrangement. To mimic this structure, \cite{Moller2016} introduced a model called Poisson line cluster point process (PLCPP), where the points are clustered around the lines of a Poisson line process. If the directional distribution of this process differs from the uniform distribution on the sphere, the resulting point process is anisotropic. See Figure \ref{fig:examplesClustered3D} for a sample of the minicolumn data and a realisation of a PLCPP model. 

\section{Analysis based on point process summary statistics}
\label{sec:SummaryStatistics}
In this and the two following sections we will review non-parametric methods for anisotropy analysis. In order to understand how typical results of an analysis may look like, we apply the methods to the regular point pattern shown in Figure \ref{fig:ExamplesGeometricAnisotropyRegular}, left, as well as the clustered pattern shown in Figure \ref{fig:examplesClustered2D}, middle. The former is an example of geometric anisotropy and is obtained from an isotropic realization of a 2D Strauss process with range of interpoint interaction $R=0.1$, strength of repulsive interaction $\gamma=0.1$, and first order parameter $\beta=100$. We then use the geometric anisotropy mechanism to compress in $y$-direction by a factor of 0.6 and stretch in $x$-direction by 1/0.6, i.e. $C=diag(1/0.6, 0.6)$. Subsequently, the pattern is rotated clockwise $\pi/6$ radians. The analysis is based on a pattern observed inside the window $W=[-1,1] \times [-1,1]$. 
The clustered pattern is an example of clustering along directed lines and is obtained as a superposition of a stationary Poisson process with intensity $\lambda=200$ with three parallel stripes formed by Poisson processes whose intensity functions are $\lambda_l(x)=100 \mathcal{N}(|x|_l; 0, 0.03^2), l=1,2,3$, where $\mathcal{N}$ stands for 1D Gaussian density function, and $|\cdot|_l$ is the distance from line $l=1,2,3$. The lines, and therefore the stripes, form a fixed angle of $\pi/5$ w.r.t. the $y$-axis.

Classical summary statistics from spatial point process theory such as the nearest neighbour distance distribution function and Ripley's K-function were originally defined for isotropic point processes. Several directional versions of these summary statistics have been formulated and are suitable for detecting anisotropies. In a typical application, a summary statistic is estimated separately for different directions, and differences between the estimates for these directions indicate anisotropy of the point pattern. Below, we discuss such analyses based on the nearest neighbour distance distribution function, Ripley's $K$ function and the pair correlation function.

\subsection{Visualization of anisotropy: Fry plot}

\begin{figure}[h!]
	\begin{center}
	\includegraphics[height=0.4 \textwidth]{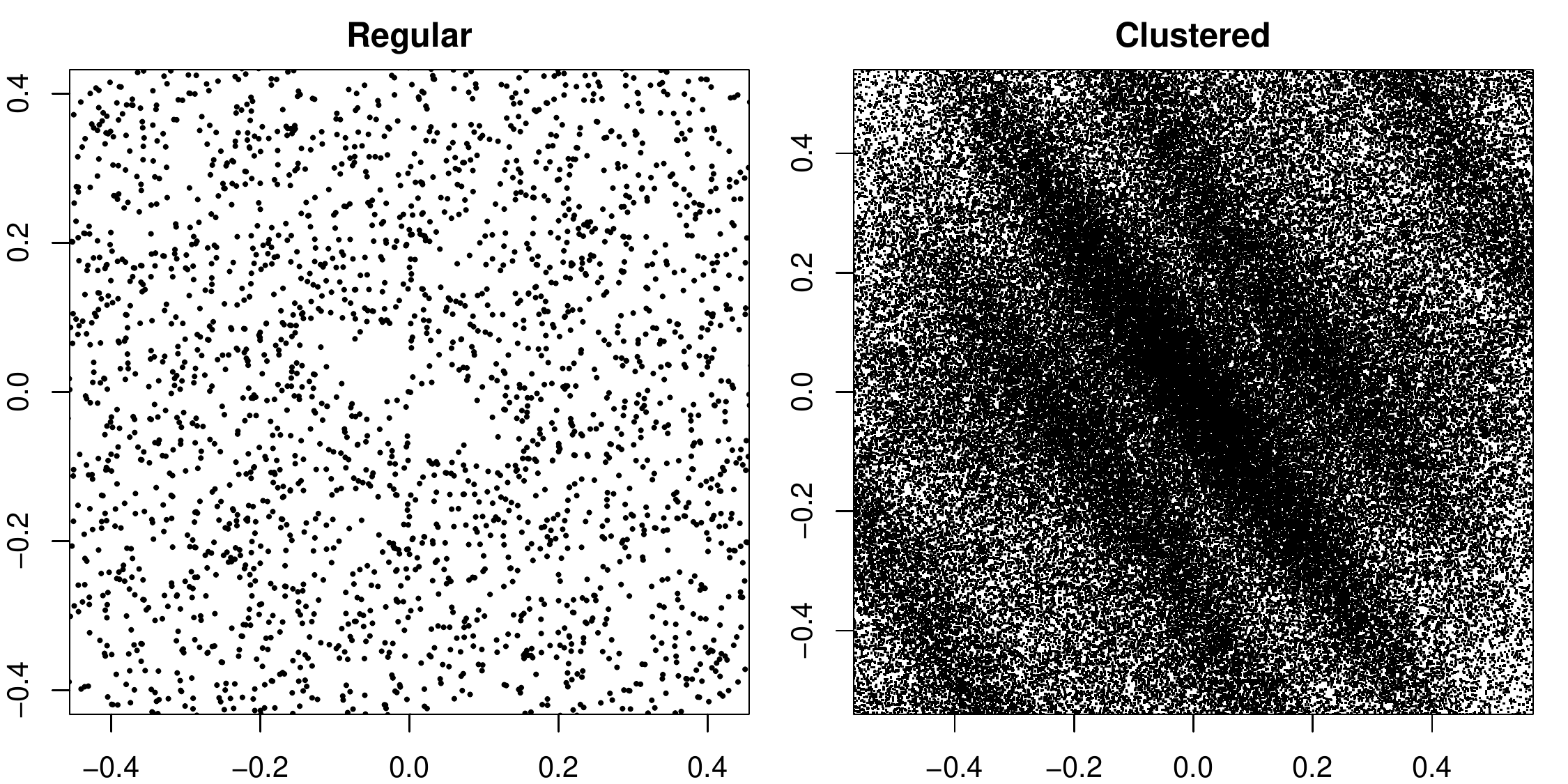}
	\end{center}
	\caption{Fry points of the example patterns shown in Figures \ref{fig:ExamplesGeometricAnisotropyRegular} and \ref{fig:examplesClustered2D} : Regular (left) and clustered (right).  }
	\label{fig:exampleFry}
\end{figure}

In clustered patterns, the shape and orientation of the clusters may reveal some anisotropies. To detect anisotropy in a regular pattern can be harder, see the point pattern in Figure \ref{fig:ExamplesGeometricAnisotropyRegular}, left. However, if we plot the pairwise difference vectors $x_i − x_j$ for all point pairs, as can be seen in Figure \ref{fig:exampleFry}, left, we can see that the central area around the origin has fewer points, indicating less pairs at short distances than long distances, typical for a regular process. We can also see that the shape of the central area is elliptic, not circular. The pairwise difference vectors do not seem to have a rotationally invariant distribution, and therefore the pattern could be anisotropic. The clustered pattern shows a somewhat inverse structure: the central area is dense, typical for clustered processes (Figure \ref{fig:exampleFry}, right). The elongated shapes taper off at the ends as we observe fewer and fewer long distance pairs due to a finite observation window.

Such plots, called \emph{Fry plots} \citep{Fry1979}, are not only visually informative but can also be very helpful in the formal analysis  as can be seen in \cite{Raj16}, who fitted ellipsoids to the Fry plot to estimate the direction of the linear transformation in the case of geometric anisotropy (see Section \ref{Sec:ellipsoid}). Nearest neighbor analysis can be based on a similar plot, where only the nearest neighbor vectors have been plotted. The Fry plot was, in fact, an improved version of plots based on the nearest neighbor distance vectors which were first introduced by \cite{Ramsay67} who applied them to measure bulk strain in rocks with rigid objects. These plots are limited, however, to 2D only: In 3D it is hard to see any structure inside the point cloud.

\subsection{Nearest neighbour analysis}
\label{subsec:nearestneighbour}

For $x \in \x$, let $\eta(x;\x) = \{y\in \x\setminus x: ||x-y||\le ||x-z||\ \forall z\in \x\setminus x\}$ denote the nearest neighbour of $x$ in the set $\x$. Let $d_i=d_i(\x)=||x_i-\eta(x_i;\x)||$ be the distance from $x_i \in \x$ to its nearest neighbour in $\x$, and let $\alpha_i=\alpha_i(\x)=\alpha(x_i-\eta(x_i;\x))$ be the angle(s) in the polar (spherical) representation of the vector from $x_i$ to its nearest neighbour in $\x$. Recall that in 2D, we have $\alpha_i= \phi_i$ while in 3D, $\alpha_i= (\phi_i,\theta_i)$.

For an isotropic point process $X$, the nearest neighbour angles are uniformly distributed i.e. $\phi \sim U[0,2\pi]$ in both 2D and 3D. In 3D, the additional angle $\theta$ is independent of $\phi$ with $\theta\sim \cos^{-1}(Z)$, $Z \sim U[-1,1].$
The corresponding angle distribution functions are given by 
\begin{equation}
F(\alpha)=F(\phi)
=\phi/2\pi, \quad \phi\in[0,2\pi] 
\label{eq:uniangle1}
\end{equation} 
and
\begin{equation}
F(\alpha)=F(\phi, \theta)= \phi(1-\cos \theta)/4\pi,\quad  (\phi, \theta)\in[0,2\pi]\times[0,\pi].
\label{eq:uniangle2}
\end{equation}

When testing isotropy via a uniformity test of the nearest neighbour angles (see Section \ref{sec:Testing}), it should be noted that the nearest neighbour angles of an observed pattern are not independent. For example, when $x_j$ is the nearest neighbour to $x_i$ and vice versa, then $\alpha_i=\check{\alpha}_j$, where $\check{\alpha}$ denotes the antipodal direction of $\alpha$. 

\cite{Illian2008a}, Ch. 4.5.2. discuss the \emph{nearest neighbour orientation density} $\varphi(\alpha)$, which is the density of the CDF $F(\alpha)$ in (\ref{eq:uniangle1}). Its edge corrected kernel estimator 
with some kernel $k_h$ and some bandwidth $h$ is given by
$$
\hat\varphi(\alpha) = \sum_{i=1}^n\frac{1(d_i < e_i)k_h(\alpha- \alpha_i)}{|W\ominus b(0, d_i)|}/\hat{\lambda}_{nn}.
$$
The edge correction reduces bias by considering only those points $x_i$ whose distance $e_i$ to the border of the observation window $W$ is larger than $d_i$. This way, only points whose nearest neighbour is observed within $W$ are included in the sampling. The intensity estimator $\hat{\lambda}_{nn}$ is adapted to this particular edge correction and is given by
$$
\hat{\lambda}_{nn}= \sum_{i=1}^n \frac{1(d_i < e_i)}{|W\ominus b(o, d_i)|.}
$$
Note that the original definition of $\varphi(\alpha)$ was for $\alpha=\phi\in[0,\pi]$ with antipodal flip when $\phi_i>\pi$. Furthermore, the kernel should be wrapped around the angle domain to avoid discontinuities.

\cite{Konig1992} define a summary for the directional distribution of the so-called $s$-nearest neighbour, i.e.\ the nearest neighbour outside some range $s$. For simplicity, we will restrict attention to the case $s=0$. Let $d_o$ and $\alpha_o$ denote the random distance and direction from the typical point $o$ of $X$ to its nearest neighbour, and consider the direction set $\A \subset S^{d-1}$.
Then the \textit{nearest neighbour directional distribution} is defined as the distribution on the unit sphere
\begin{equation}
D_{r}(\A)= 
P_o(\alpha_o\in \A|d_o < r), \quad \A \subset S^{d-1}.
\label{eq:dirdistr}
\end{equation}
Consequently, $D_{r}(\A)$ can be interpreted as the probability that the nearest neighbour of the typical point of $X$ is in $\A$ given that the distance to the nearest neighbour is at most $r$.
A consistent and asymptotically unbiased estimator for the directional distribution 
is given by
\begin{equation}
\hat{D}_{r}(\A) 
= \frac{\sum_{i=1}^n1(d_i < r)1(\alpha_i\in \A)1(x_i\in W\ominus b(o,r))}{\sum_{i=1}^n 1(d_i <r)1(x_i\in W\ominus b(o,r))}. 
\label{eq:dirdistrEst}
\end{equation}


\cite{Redenbach2009} consider a similar statistic where the roles of $\A$ and $r$ are exchanged. The \emph{global directional nearest neighbour distance distribution} function $G_{\glob,\A}$ is defined as the distribution of $d_o$ conditioned on $\alpha_o \in \A$,
$$
G_{\glob,\A}(r)=P_o(d_o < r | \alpha_o\in \A).
$$
An estimator for this statistic is given by
$$
\hat{G}_{\glob,\A}(r)= \frac{\sum_{i=1}^n 1(d_i < r)1(\alpha_i\in \A)1(x_i\in W\ominus b(o,d_i))}{\sum_{i=1}^n 1(\alpha_i\in \A) 1(x_i\in W \ominus b(0, d_i))}.
$$

In \cite{Redenbach2009}, the sets $\A$ are chosen such that the angle between elements of $\A$ and a prespecified direction $\alpha$ is less than a given $\epsilon$. Hence, only points with the nearest neighbour in a double cone $C(\alpha, \epsilon)$ are considered in the estimation, which may drastically reduce the sample size. As an alternative statistic, \cite{Redenbach2009} also consider a local nearest neighbour distance distribution function $G_{\loc,\alpha,\epsilon}$ which is the cumulative distribution function of the distance from a typical point to the nearest neighbour in the double cone $C(\alpha, \epsilon)$, or
$$
G_{\loc,\alpha,\epsilon}(r)=P_o\left[ d_o\left(X\cap C(\alpha,\epsilon)\right) < r \right].
$$

Writing $d_i^{\alpha,\epsilon}=d_i(\x\cap (x_i+C(\alpha,\epsilon)))$ for the $C(\alpha,\epsilon)$-nearest neighbour distance of $x_i$, \cite{Redenbach2009} propose the Hanisch type estimator
$$
\hat{G}_{\loc,\alpha, \epsilon}(r)= \frac{\hat{G}_{H, \loc, \alpha, \epsilon}(r)}{\hat{G}_{H, \loc, \alpha, \epsilon}(\infty)},
$$
where


$$
\hat{G}_{H,\loc, \alpha, \epsilon}(r) = \sum_{i=1}^n \frac{1(d_i^{\alpha,\epsilon} < r)1(x_i\in W \ominus  S(\alpha,\epsilon, d_i^{\alpha,\epsilon}))}{|W \ominus S(\alpha,\epsilon, d_i^{\alpha,\epsilon}) |}. 
$$
This way, each point of the process contributes to the estimation of $G_{\loc}$ (unless disregarded by the edge correction). 

\begin{figure}
	\begin{center}
	\includegraphics[width=1 \textwidth]{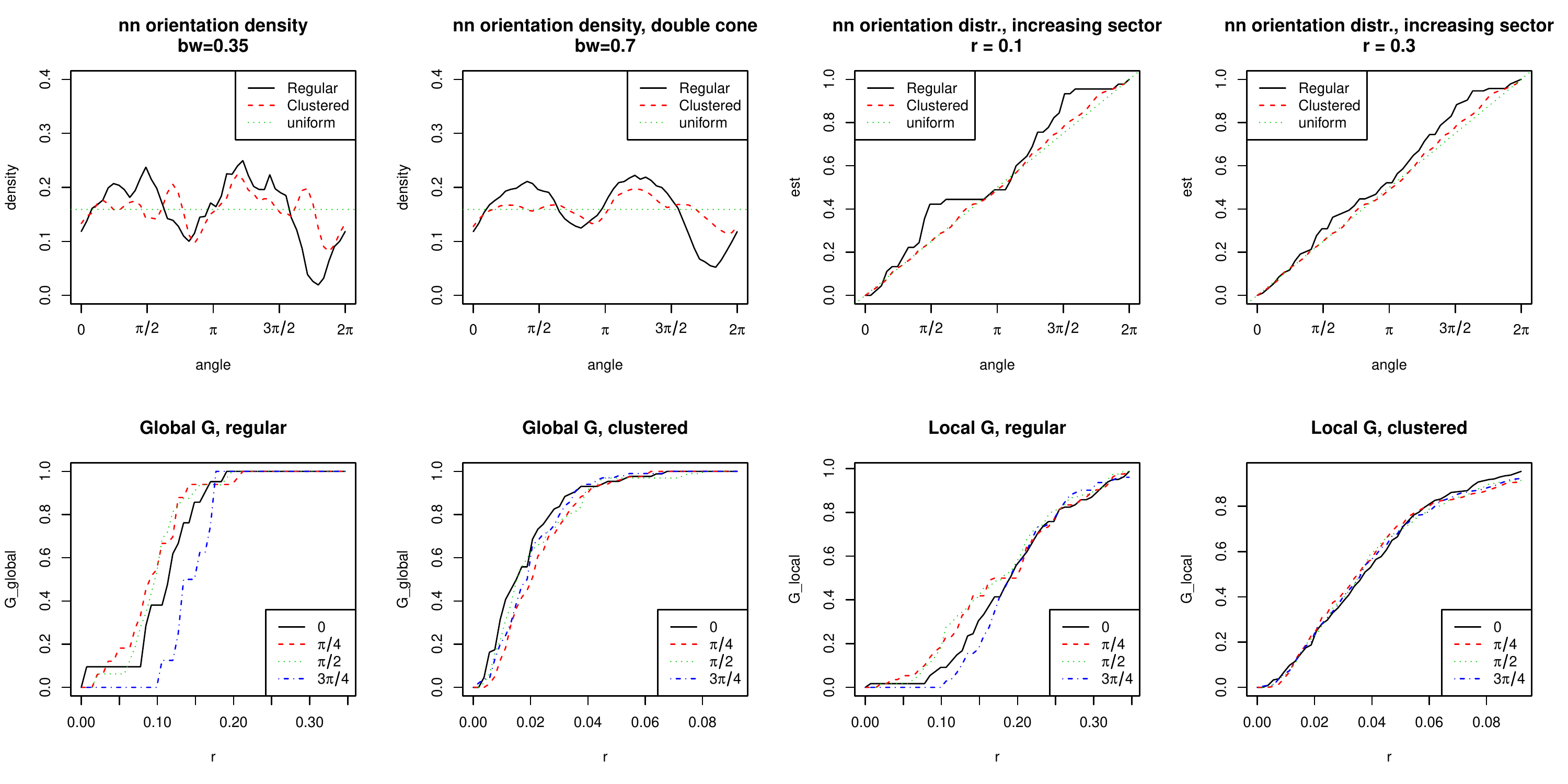}
 	\end{center}
	\caption{Directional nearest neighbour summaries for the two example patterns.}
	\label{fig:exampleNN}
\end{figure}

Figure \ref{fig:exampleNN} depicts the four nearest neighbour summaries for the regular and clustered example patterns in Figure 3 and 4. The two leftmost plots in the top row show the orientation density. No guidelines were given in \cite{Illian2008a} for the bandwidth, so the first estimate uses a bandwidth selected by eye and the second plot a bandwidth double that of the first estimate. The smoother curve for the regular pattern indicates two peaks around $\alpha=\pi/2-\pi/6$ and $\alpha+\pi$, correctly identifying the rotation in the pattern. The two rightmost plots on the top row depict the directional distribution plots, which were computed with $\A=\A(a)=\{u\in S^1: \alpha(u)\in[0,a]\}, a\in[0,2\pi]$. No guidelines were given by \cite{Konig1992} for choosing the ranges, so we chose $r=0.1$ and $r=0.3$. The former is the known interaction range for the regular process, and the latter shows the directional distribution for all points. The two peaks for the regular pattern are visible at short range $r=0.1$ but not at longer range $r=0.3$, so the anisotropy information is present only in short nearest neighbour distances. The directed $G$-functions in directions $\alpha=0,\pi/4,\pi/2$, and $3\pi/4$ are plotted in the bottom row. \cite{Redenbach2009} do not discuss the choice of the sector half-angle $\epsilon$. To guarantee that there is no overlap between the double cones, $\pi/8$ is the largest half-angle that can be chosen. Here, we chose $\pi/4$ to double the sample sizes (we have only 122 points in the regular case, of which only $\epsilon/\pi$ on average contribute per direction). The $G$-function values in directions close to the correct rotation angle differ from those computed along the coordinate axes, indicating departure from anisotropy. All summaries detect poorly the anisotropy of the clustered pattern, as the pattern behaves like an isotropic Poisson process at the very small scales where the nearest neighbour information is concentrated. 


\subsection{Second order analysis: Ripley's $K$ function and the pair-correlation function}
\label{sec:SecondOrder}

Nearest-neighbour characteristics are known to be rather short-sighted as demonstrated by the examples discussed above. Therefore, Ripley's $K$ function and the pair correlation function (pcf) may be better tools for directional analysis. There are several ways to derive a second order summary that takes direction into account. The first suggestion can be found in \cite{Ohser1981}. They define in 2D an angle dependent version of the K-function, such that the function is cumulative in both range and angle.

\subsubsection{K-functions}
\label{subsec:K}

Second order statistics are based on the second-order factorial moment measure
\[
\alpha^{(2)}(A\times B)=\E\left( \sum_{x,y \in X}^{\neq} 1(x\in A)1(y\in B) \right)
\]
for Borel sets $A$ and $B$ in $\R^d$, where the upper $\neq$ means that only pairs with $x\neq y$ are considered in the sum.
We assume that the (second order) product density $\rho^{(2)}$ exists, i.e.
\[
\alpha^{(2)}(A\times B) = \int_A\int_B \rho^{(2)}(x,y)dydx.
\]
The value $\rho^{(2)}(x,y)dxdy$ is the probability that $X$ has a point in each of the infinitesimally small discs with centers $x$ and $y$ and volumes $dx$ and $dy$, respectively. 
For stationary $X$, we have that $\rho^{(2)}(x,y) = \rho^{(2)}(o,y-x)=:\rho^{(2)}(z)$ where $z=y-x$. Hence,
$$
\alpha^{(2)}(A\times B)= \int_A\int_B \rho^{(2)}(o,y-x)dydx = \int_A\int_{B-x} \rho^{(2)}(z)dzdx.
$$
Defining the reduced second-order moment measure $\K$ via 
$$
\lambda^2\K(B) = \int_B \rho^{(2)}(z) dz,
$$
we arrive at
$$
\alpha^{(2)}(A\times B) = \lambda^2\int_A \K(B-x)dx.
$$
Alternatively,  by the Campbell-Mecke formula  \citep[][Eq.(4.1.8)]{Illian2008a}
$$
\alpha^{(2)}(A\times B)=\E \sum_{x\in X}1(x\in A) X(B\setminus\{x\})=\lambda\int_A \E_o[ X((B-x)\setminus\{o\})]dx.
$$
These two equations lead to $$\lambda \K(B) = \E_o( X(B \setminus \{o\})),$$ and we can interpret $\lambda\K(B)$ as the expected number of (further) points in $B$ conditioned on $o \in X$.  

In the case of geometric anisotropy, i.e.\ when we consider $X=TX_0$, where $T$ is an invertible linear mapping and $X_0$ a stationary and isotropic point process, the point process $X$ has the following further property
\begin{enumerate}
\item[5.]
The reduced second order moment measures are related by $\K_X(B)=\det(T) \K_{X_0}(T^{-1}B)$ for any Borel set $B \subset \R^d$.
\end{enumerate}

An estimator for $\lambda^2 \K(B)$ is given by
$$
\widehat{ \lambda^2\K}(B)= \sum_{x,y \in \x}^{\neq} \frac{1(y-x\in B)}{|W_x \cap W_y|}.
$$
Unbiasedness of the estimator follows from the following generalization of the Campbell theorem
\begin{equation}
\label{eq:Campbell}
\E \sum_{x,y \in X}^{\neq}f(x,y) =\int\int f(x,y)\alpha^{(2)}(d(x,y))=\lambda^2\int\int f(x, x+h)dx \K(dh)
\end{equation}
for all nicely behaving $f$, see \citet[p.\ 228]{Illian2008a}. 
The choice $K(r)= \K(b(o,r))$ yields Ripley's K-function.  Note that $\rho^{(2)}$ is commutative, so all summaries based  on the factorial moment measure are antipodally symmetric such that only the upper hemisphere of directions in $S^{d-1}$ needs to be considered.

\begin{figure}[h!]
	\begin{center}
	    \includegraphics[width=0.5 \paperwidth]{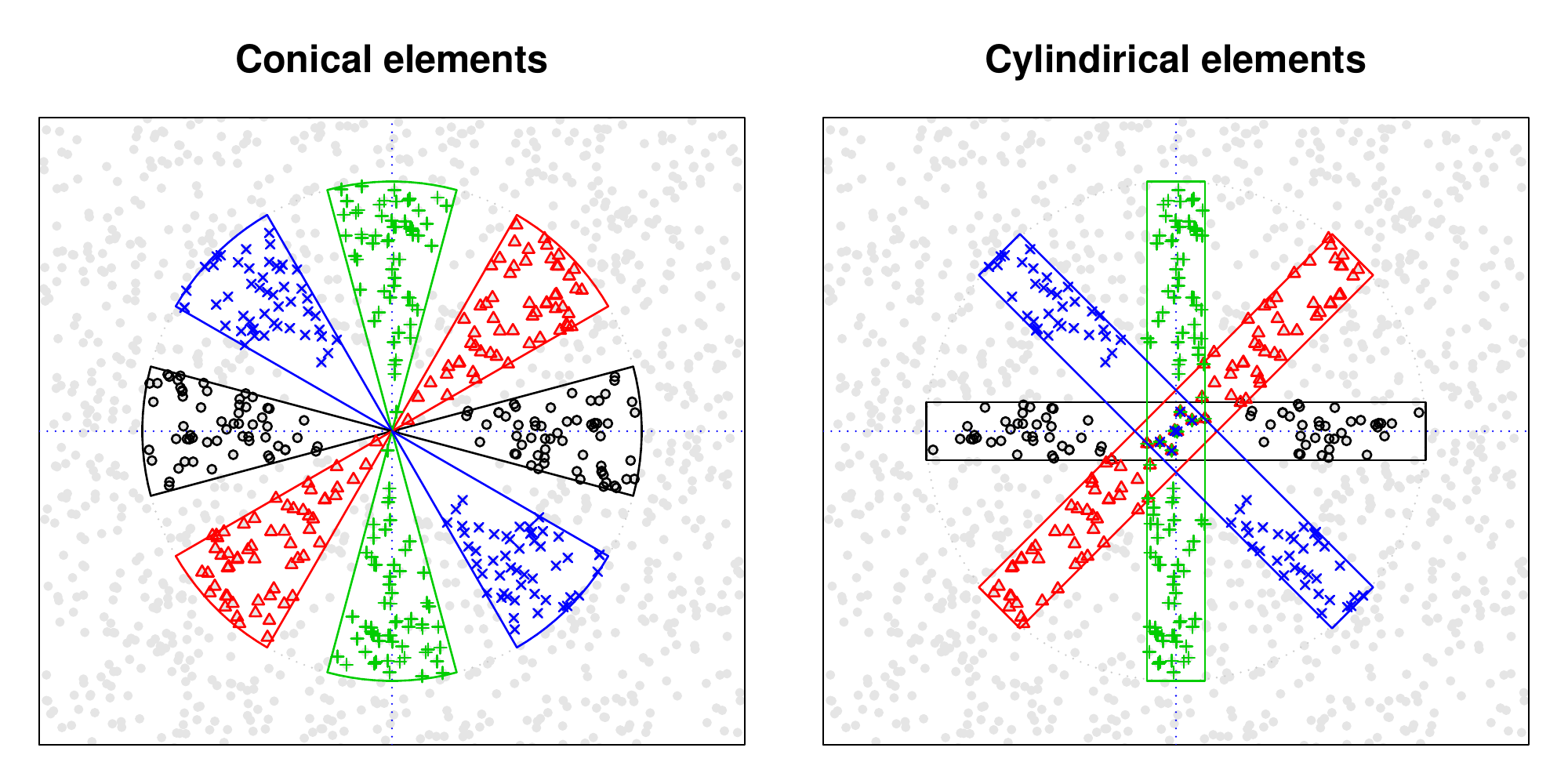}
	\end{center}
	\caption{The anisotropic test sets for the conical (left) and cylindrical (right) $K$ functions, overlaid on top of the Fry plot of the regular example pattern. }
	\label{fig:K-test-sets-2d}
\end{figure}

Anisotropic second order summaries can now be defined in terms of parametric, rotation variant test-sets $B=B(\psi)$, where $\psi$ parameterizes the set in terms of direction and length-scale. An early 2D example of such a construction was given by \cite{Ohser1981}, who chose $B$ to be the sector of radius $r$ with angle $\gamma$ from the positive $x$-axis and considered this as a function $K(r, \gamma)$. This has an interpretation as a cumulant in $\gamma\in[0,2\pi]$. \cite{Stoyan1991} then extended this idea to a more freely defined sector, where both bounding rays can have arbitrary angles $\gamma$ and $\Gamma> \gamma$ w.r.t.\ the $x$-axis. 
Setting $\alpha= (\gamma+\Gamma)/2$ and $\epsilon= (\Gamma-\gamma)/2$ and using the sector $S(\alpha,\epsilon,r)$ as the test set, the 3D extension becomes obvious. Figure \ref{fig:K-test-sets-2d}, left panel, illustrates the construction in 2D. The resulting version of the $K$-function is called the \emph{conical $K$-function}, and was used by \cite{Redenbach2009} to assess anisotropy in 3D patterns by contrasting the $z$-direction to $x$- and $y$-directions.

An alternative choice is the \emph{cylindrical $K$-function} \citep{Moller2016}. The test-set is the origin centred rectangle (2D) or cylinder (3D) $L(r, u, h_c)$ with major-axial direction unit vector $u\in S^{d-1}$, with height $2r$ and cross-section half-length $h_c, 0<h_c<r$, i.e. $L(r,u,h_c)=\{x\in \R^d: d(x, l(u)) \le h_c, d(x, l(u)^{\perp}) \le r\}$, where $l(u)$ is the line spanned by $u$, $l(u)^{\perp}$ is the orthogonal hyperplane, and $d(x, l(u))$ and $d(x, l(u)^{\perp})$ denote the distances from a point $x$ to the line and the hyperplane, respectively.
Figure \ref{fig:K-test-sets-2d}, right panel, illustrates this construction. Two notable differences to the conical case are that the local area of the test set is not increasing in $r$, and that the test-sets in different directions always overlap at short ranges.

\cite{Illian2008a} Ch.\ 4.5.3.\ define an orientation distribution in 2D using another version of the $K$ function. Let $T(\alpha,r)$ be the $r$-sector formed by the positive $x$-axis and angle $\alpha$, and let $T(\alpha, r_1,r_2) = T(\alpha,r_2)\setminus T(\alpha,r_1)$ for $r_1<r_2$.  Then we set $K(r_{1}, r_2, \alpha)=\K(T(\alpha, r_1,r_2))$. For fixed ranges $r_{1}$ and $r_{2}$,
$$F_K(\alpha) = \frac{K(r_1, r_2, \alpha)}{K(r_1, r_2, \pi)}$$
is an orientation distribution function with a density $f_{r_1,r_2}(\alpha)$, which we call \emph{second order orientation density}. Its estimator is, up to the constant $K(r_1, r_2, \pi)$,
\[
\hat f_{r_1,r_2}(\alpha) \propto \sum_{x,y \in \x}^{\neq} \frac{1(r_1<||x-y||<r_2)k_{h_\alpha}(\alpha(x,y)-\alpha)}{|W_x \cap W_y |}
\]
with some  kernel $k_{h_\alpha}$ and bandwidth $h_\alpha$. For isotropic processes $f_{r_1,r_2}\equiv 1/\pi$.
The definition can clearly be generalized to 3D. However, its application is complicated since in this case a function of two angles has to be considered.

\begin{figure}
	\begin{center}
	    \includegraphics[width=0.95 \textwidth]{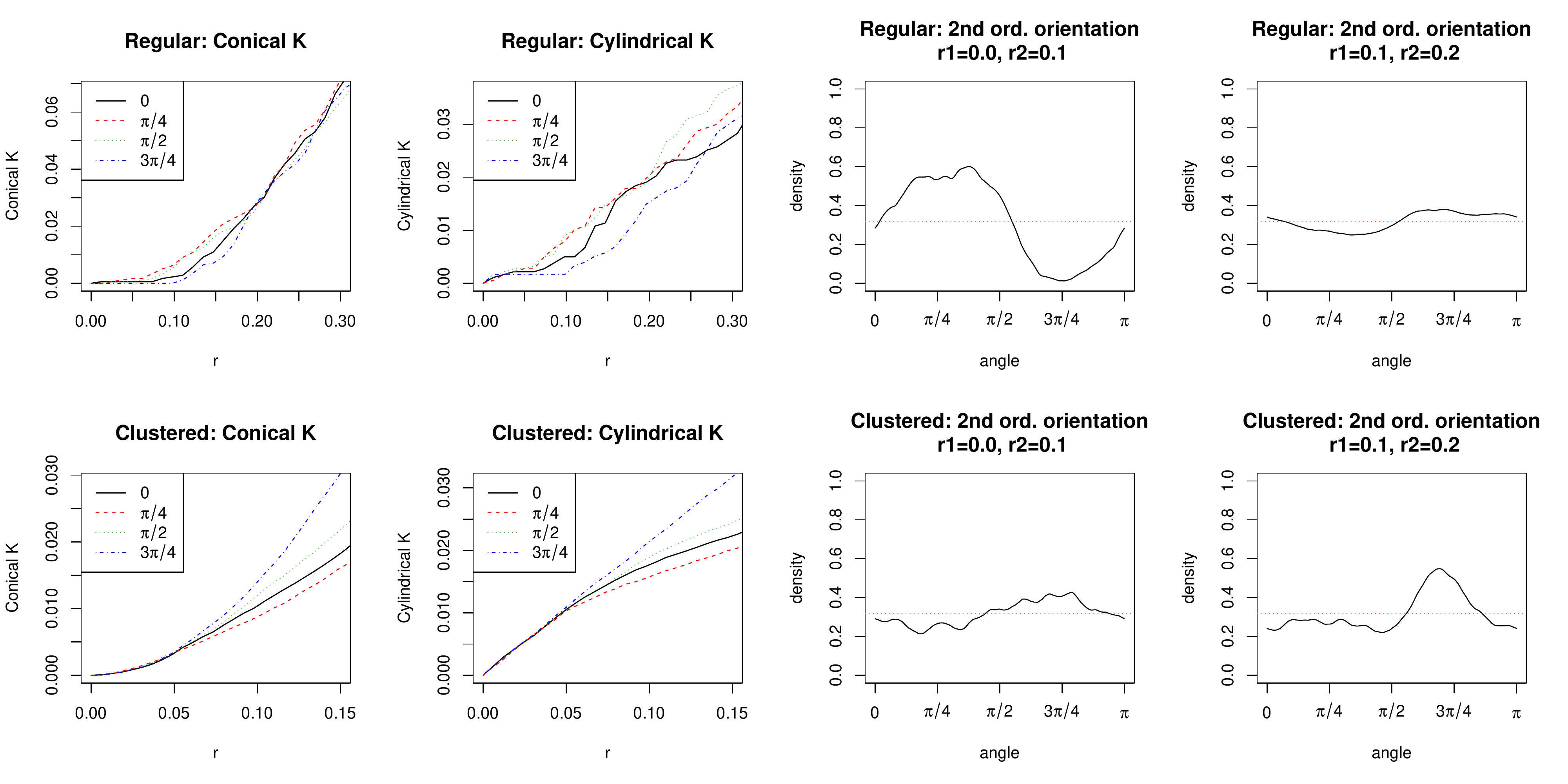}
	\end{center}
	\caption{The conical $K$ with $\epsilon=\pi/8$, cylindrical $K$ with $h_c=0.03$, and orientation densities with $(r_1=0, r_2=0.1)$ and $(r_1=0.1, r_2=0.2)$ and bandwidth $h_\alpha=3/\sqrt{\lambda}$.  Top row: Regular pattern. Bottom row: Clustered pattern.}
	\label{fig:K-summaries-example} 
\end{figure}

Figure \ref{fig:K-summaries-example} depicts the second order summaries for the example patterns in directions  $\alpha=0,\pi/4,\pi/2,3\pi/4$. For the conical $K$, no guidelines for selecting the sector half-angle are available, so we chose the maximal angle with no overlaps, $\pi/8$. Guidelines for the cylinder cross-section half-lengths under specific volume and shape constraints are discussed by \cite{Safavimanesh2016}, who derive equations connecting the cylinder height and half-length to a corresponding conical shape. For simplicity, we fixed $h_c=0.03$ by eye (as shown in Figure \ref{fig:K-test-sets-2d}). The conical and cylindrical $K$-functions show decreased amount of pairwise directions near the true stretching direction and increased amount near the direction of compression of the regular pattern when compared to alternative directions. In the clustered pattern, there are most pairs in the direction of the clusters. For the orientation density, in \cite{Illian2008a} the range interval $[r_1,r_2]$ was "found by experimentation". For the example, we selected two intervals corresponding to "short" and "long" ranges. The level of smoothing was not discussed, so by some trial and error we decided on $h_\alpha=3/\sqrt{\lambda}$ for the Epanechnikov kernel, leading to stable looking estimates. The orientation density captures the anisotropy of the regular process well when using a short range interval $[r_1,r_2]$ but is practically uniform for long ranges, and vice versa for the clustered pattern.  

\begin{figure}
	\begin{center}
	    \includegraphics[width=0.6 \textwidth]{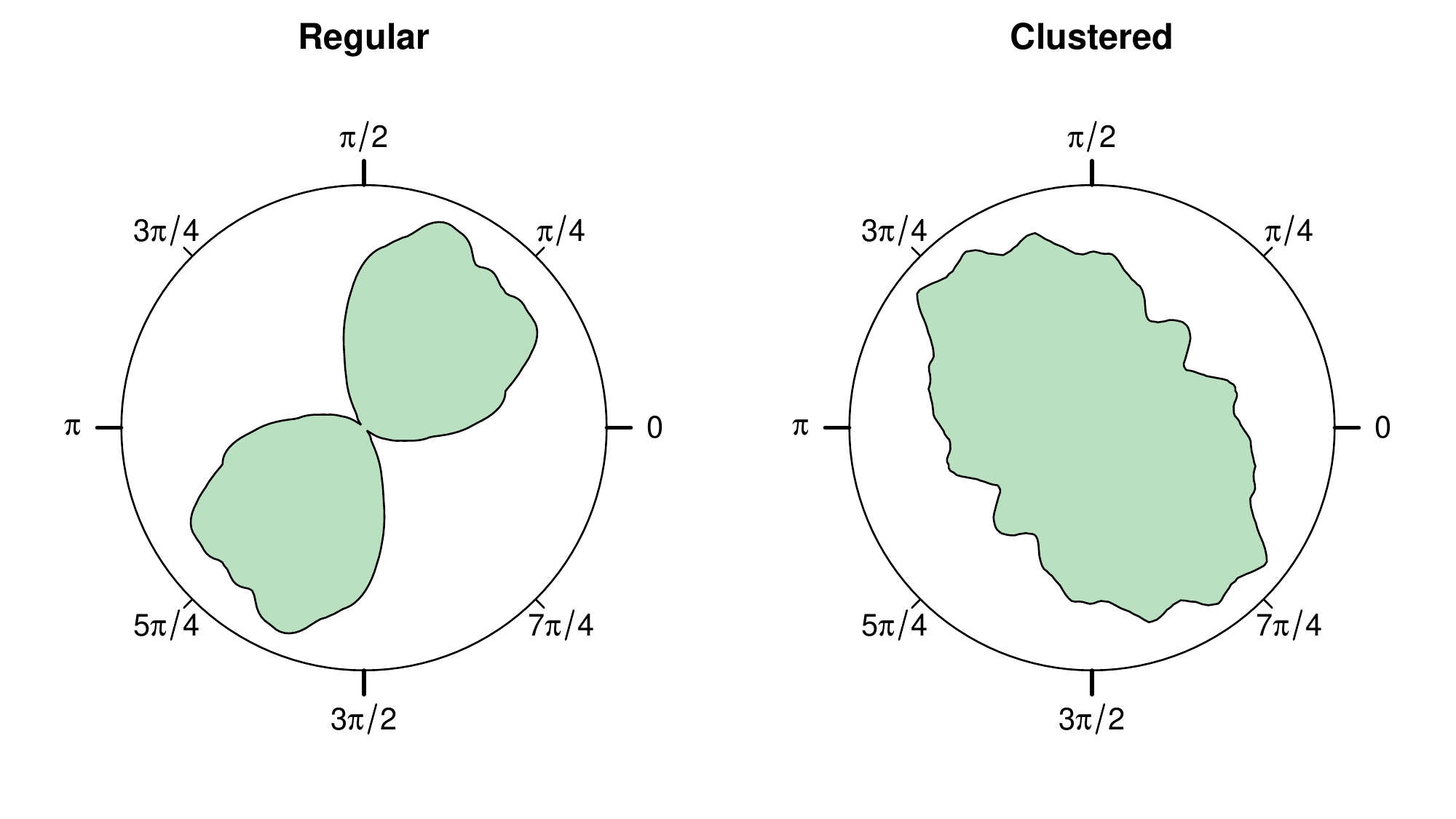}
	\end{center}
	\caption{The rose-of-directions plots of the 2nd order orientation densities in Figure \ref{fig:K-summaries-example} with $r_1=0$ and $r_2=0.1$.}
	\label{fig:fry-rose}
\end{figure}

An alternative visualisation of any density function over angles in 2D is given by the rose-of-directions \citep{MeckeStoyan80}. It depicts the density over a rotation, emphasizing departures from a circle. Figure \ref{fig:fry-rose} illustrates the rose for the short range 2nd order orientation densities of our examples. 

All summary statistics introduced in this section will most clearly reveal the anisotropy if $r$ is chosen according to relevant scales in the point pattern. For instance, the hardcore or cluster radius $R$ of geometric anisotropic point processes may strongly vary in different directions. Hence, obvious differences in the directional K-function can be expected for $r \approx R$. 

\subsubsection{Pair correlation functions (pcf)}
\label{subsec:pcf}
The cumulative nature of the $\K$-measures can sometimes obscure fine details, and their derivatives offer a more detailed option. We can rewrite the definition of the $\K$-measure as 
$$
\K(B)=\lambda^{-2}\int_B\rho^{(2)}(z)dz = \int_B g(z) dz,
$$
where the integrand
$$g(z) = \frac{\rho^{(2)}(z)}{\lambda^2}$$
is called the pair correlation function (pcf). The pcf is more practical than the product density as it is independent of intensity and the Poisson process has $g\equiv 1$. Note that it is, however, not a correlation in the conventional statistical sense as it takes values on $[0,\infty)$. In the case of geometric anisotropy, the process $X$ has the following further properties 
\begin{itemize}
\item[6.]
The second order product densities (if they exist) fulfill $\rho_X^{(2)}(z)=\det(T)^{-2} \rho_{X_0}^{(2)}(T^{-1}z)$ for all $z\in \R^d$.
\item[7.]
The pair correlation functions (if they exist) are related via $g_X(z)=g_{X_0}(T^{-1}z)$ for all $z\in \R^d$.
\end{itemize}

Given a parametric set $B=B(\psi)$ for the $\K$-measure, we then define the anisotropic pair correlation function as the Radon-Nikodym derivative
$$\mathcal{K}(B(\psi)) = \int_{B(\psi)} g(y) dy.$$
For a set $B$, where the integral has a simple geometrical decomposition, the corresponding derivatives have simple interpretations as anisotropic pcf's. For example, the conical set gives
$$\K(S(u, \epsilon, r)) = \int_0^r\int_{v\in S^{d-1}: cos^{-1}(v\cdot u)<\epsilon} g(v,t) dv dt$$
with $g(v,t) = g(vt)$ for unit vectors $v\in S^{d-1}$. The change to polar coordinates in 2D leads to the simple form 
$$\K(S(u, \epsilon, r)) = \K(S(\alpha(u), \epsilon, r))=  2 \int_0^r\int_{\alpha(u)-\epsilon}^{\alpha+\epsilon}g(a,t) t\ da dt,$$
where $g(a,t) = g(u,t)$ for direction $u$ with $\alpha(u)=a$. The expression is multiplied by 2 as $S(u, \epsilon, r)$ is defined as a double cone. The cylindrical case is more complicated.

A simplification of the three parameter case is to assume that $g(v,t)$ is constant over the arc of the $r$-sector or the rotation cap of the $r$-cone. Then for a fixed $\epsilon$ 
$$\K(S(u, \epsilon, r)) = \int_0^r g_u(t) dt$$
with 
$$\frac{g_u(t)}{v(\epsilon,r)} = g(ut),$$ 
where the length or area $v(\epsilon,r)$ of the cap is $2r\epsilon$ when $d=2$ and $2\pi r^2(1-cos(\epsilon))$ when $d=3$. We call the function $g_u(t)$ the \emph{conical pcf} \citep{Stoyan1991}. Note that assuming that the pcf is isotropic is the same as setting $\epsilon = \pi$.

For the cylinder element $L(r, u, h_c)$ we can also simplify the corresponding pcf by assuming that for each $r$ the $g$ is approximately constant over the cylinder cross section, and then, by holding the direction $u$ and the half-width of the cylinder $h$ fixed, define the \emph{cylindrical pcf} as a function of range for which
$$\K(L(r,u,h_c)) = \int_0^r g^c_u(t)dt$$
so that 
$$\frac{g^c_u(t)}{b_{d-1}h_c^{d-1}} = g(ut) \quad\text{for } ut\in \{x: ||x - (x\cdot u)u||<h_c\},$$
where $b_d$ is the volume of $d$-dimensional unit ball.





The estimator for the anisotropic pcf in 2D as given by \cite{Stoyan1991} and also replicated by \cite{Moller2014} is
$$\hat{g}(a,r)= \frac{1}{4r\hat{\lambda}^2}\sum_{x,y \in \x}^{\neq} \frac{w_h(y-x, (r,a))+w_h(y-x, (r,\check{a}))}{|W_x \cap W_y |}$$
with intensity estimator $\hat\lambda$, kernel function $$w_h(y-x, (r,a))=k_{h_r}(||y-x||-r)k_{h_a}(\alpha(y-x)-a),$$ smoothing parameters $h=(h_r,h_a)$, and the antipodal direction $\check{a}$ of $a$. For 3D with spherical coordinates $(r, \phi, \theta)$, the analogous estimator is 
$$\hat{g}(\phi, \theta, r) = \frac{1}{4 r^2 \sin(\theta) \hat{\lambda}^2}\sum_{x,y \in \x}^{\neq} \frac{w_h(y-x, (r,\phi,\theta))+w_h(y-x, (r,\check{\phi},\check{\theta}))}{|W_x \cap W_y |}$$
with an additional kernel element for the second angle. Note that the estimator behaves poorly for $\theta$ close to 0 or $\pi$, which corresponds to the unit vectors (0,0,$\pm$1). 


An alternative estimator for the anisotropic pcf is given by
\[
\hat{g}(u,r) = \frac{1}{\hat{\lambda}^2}\sum_{x,y \in \x}^{\neq} \frac{k_{h_d}(x-y-ru)}{|W_x \cap W_y |}
\]
with some $\R^d$ kernel $k_{h_d}$ \citep{Guan2007}.  This estimator does not suffer from the small denominator problem, but it introduces bias near the origin as the $\R^d$ smoothing kernels  for different directions might overlap at small $r$. 

The conical pcf with central half-angle $\epsilon$ and direction $u$ can be estimated by
$$\hat{g}_u(r) = \frac{1}{v(\epsilon,r)\hat{\lambda}^2}\sum_{x,y \in \x}^{\neq} \frac{1(\alpha(y-x,u)<\epsilon)k_{h_r}(||y-x||-r)}{|W_x \cap W_y |}$$
with $\alpha(v,u)=acos(\frac{v\cdot u}{||v|||u||})$. Note that the conical estimator is the same as the anisotropic estimator with a box kernel for the angles. Similarly, the cylindrical pcf with a cylinder half-height $r$, half-diameter $h_c$ and direction $u$ can be estimated by
$$\hat{g}^c_u(r) = \frac{1}{b_{d-1}h_c^{d-1}\hat{\lambda}^2}\sum_{x,y \in \x}^{\neq} \frac{1(||(y-x) - [(y-x)\cdot u]u||<h_c)k_{h_r}(|(y-x)\cdot u|-r)}{|W_x \cap W_y |}.$$


\begin{figure}
	\begin{center}
	    \includegraphics[width=0.95 \textwidth]{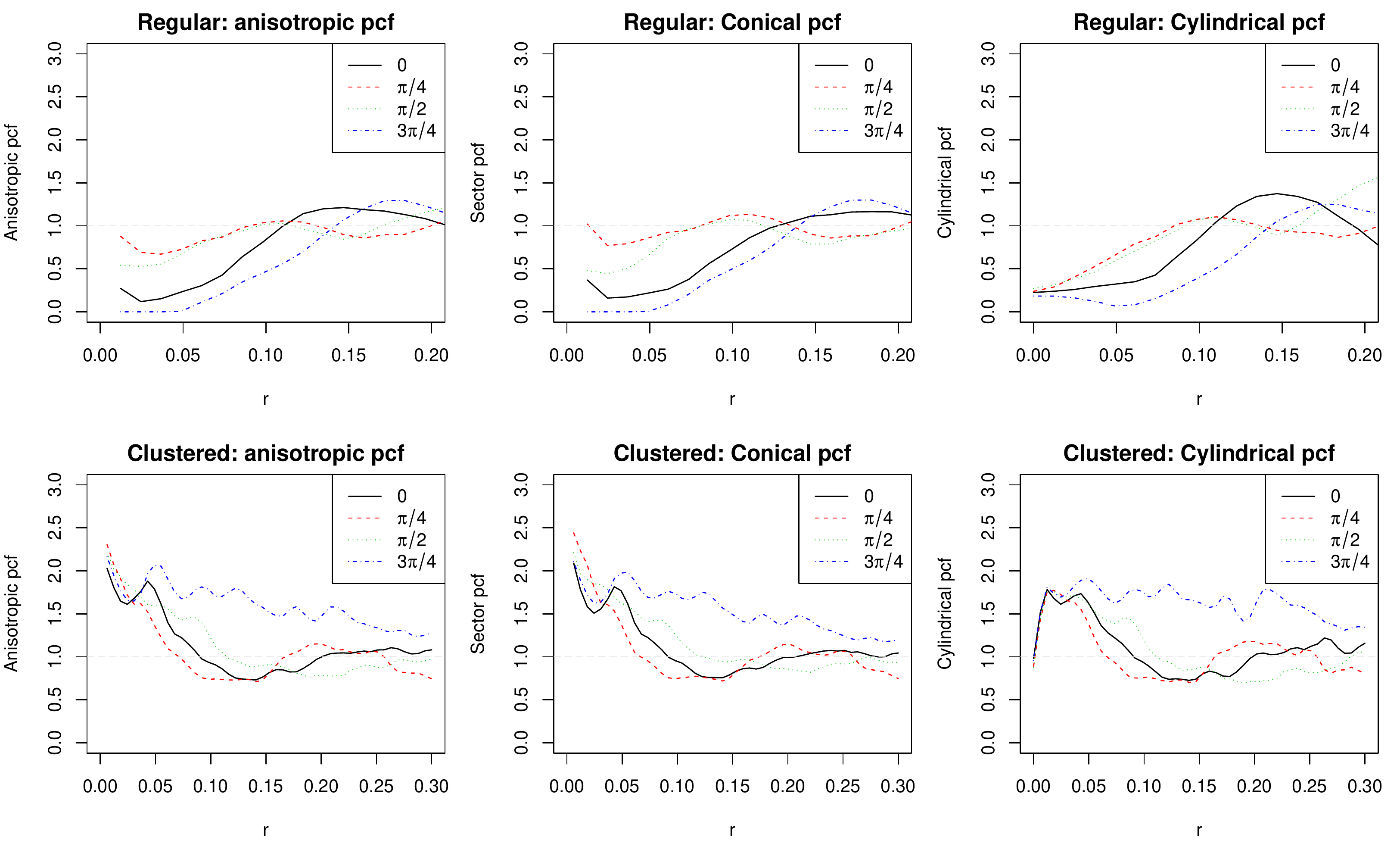}
	\end{center}
	\caption{The anisotropic, conical and cylindrical pcf estimates for the example patterns, with $h_\alpha=\epsilon=\pi/8$, cylindrical width       
    $h_c=0.03$, range bandwidth $h_r=0.3/\sqrt{\lambda}$.  Top row: Regular pattern. Bottom row: Clustered pattern.}
	\label{fig:pcf-summaries-example}
\end{figure}

These estimators have generally two tuning parameters. Several methods for choosing the range smoothing bandwidth have been proposed for isotropic pcfs, with Stoyan's rule of thumb $h_r\approx 0.15/\sqrt{\lambda}$ \citep[][Sec.4.3.3]{Illian2008a} still being the most used one. \cite{Guan2006} suggest a block-sampling based optimisation approach for the alternative anisotropic pcf estimator's only bandwidth. Further guidelines for anisotropic smoothing have not been presented. 

Figure \ref{fig:pcf-summaries-example} depicts the anisotropic, conical and cylindrical pcf estimates for the two example patterns. The sector half-angle for the conical pcf and the angular smoothing bandwidth for the anisotropic pcf were set to $\pi/8$. The cylinder half-width was set to $h=0.03$, the same as for the cylindrical $K$-function in the previous section. For the range bandwidth we increased Stoyan's rule of thumb to $0.3/\sqrt{\lambda}$ to account for reduced sample sizes due to directional sub-sampling. The anisotropic  and conical pcfs are very similar, depicting differences between directions for the regular pattern at short ranges and for the clustered pattern at longer ranges. The cylindrical sets overlap at short ranges so that the estimates in different directions are similar, but the longer range differences in the clustered pattern are captured well. 

\subsubsection{Estimation of preferred directions: Fitting ellipsoids to Fry points}
\label{Sec:ellipsoid}
To detect geometric anisotropy, \cite{Raj16} suggested to fit ellipsoids (ellipses in 2D) to the contours of directed cumulants of the pairwise difference vectors, i.e.\ unscaled sector-K functions. The fitted ellipsoid not being a ball is an indication of anisotropy and the favored direction can be estimated based on the orientation of the ellipsoid. They define a set of so-called pseudo Fry points $G_l:= \{g_u=r_l(u)u: u\in U\}$, where $r_l(u)$ is the distance to the $l$th nearest Fry point in the sector $S(u,\epsilon)$ and $U$ is the set of directions of interest. 
For a fixed set $L\subset\mathbb{N}$ of contour levels $l$, one has then a collection of pseudo-Fry point sets $\{G_l: l\in L\}$, from which the contour ellipsoids $\{E_l:l \in L\}$ can be estimated. More specifically, for each $l\in L$, the observed points $g_u\in G_l$  are assumed to follow a measurement error model
$$
g_u = e_u + \varepsilon_u, \qquad u \in U,
$$  where $e_u\in E_l$ are the true contour points, fulfilling the origin centered quadratic equation 
$e_u^T A_le_u=m_l$ 
with a semi-definite matrix $A_l$ and scale parameter $m_l$, and $\varepsilon_u$'s are independent measurement errors with Gaussian distribution $\varepsilon_u\sim N(0,\sigma^2 I)$
with $I$ being the identity matrix. 
It is enough to consider origin centered ellipsoids since the underlying process is stationary. The scale parameters can be fixed to $m_l=1$. The model can be fitted using penalized and adjusted ordinary least squares to obtain $\hat A_l$ \citep{Kukush2004}. A rotation matrix estimate can then be derived using an eigenvalue decomposition of $\hat A_l$. To produce an estimate of the average rotation, \cite{Raj16} sample a set of contour ellipsoids with noise and fit another ellipsoid to the superimposed samples. 

The method is sensitive to the choice of the contour levels, the number of directions and the sector angles. The set of useful contours depends on whether the process is regular or clustered, and on the intensity. In addition, if too many narrow sectors are used, they will not capture enough Fry points and the ellipses represent poorly the contours. Too few, wide sectors, on the other hand, bias towards circular contours. Further technical issues involving border correction and the double-effect of each Fry point pair are omitted here.



\begin{figure}[h!]
	\begin{center}
        \includegraphics[width=0.6 \paperwidth]{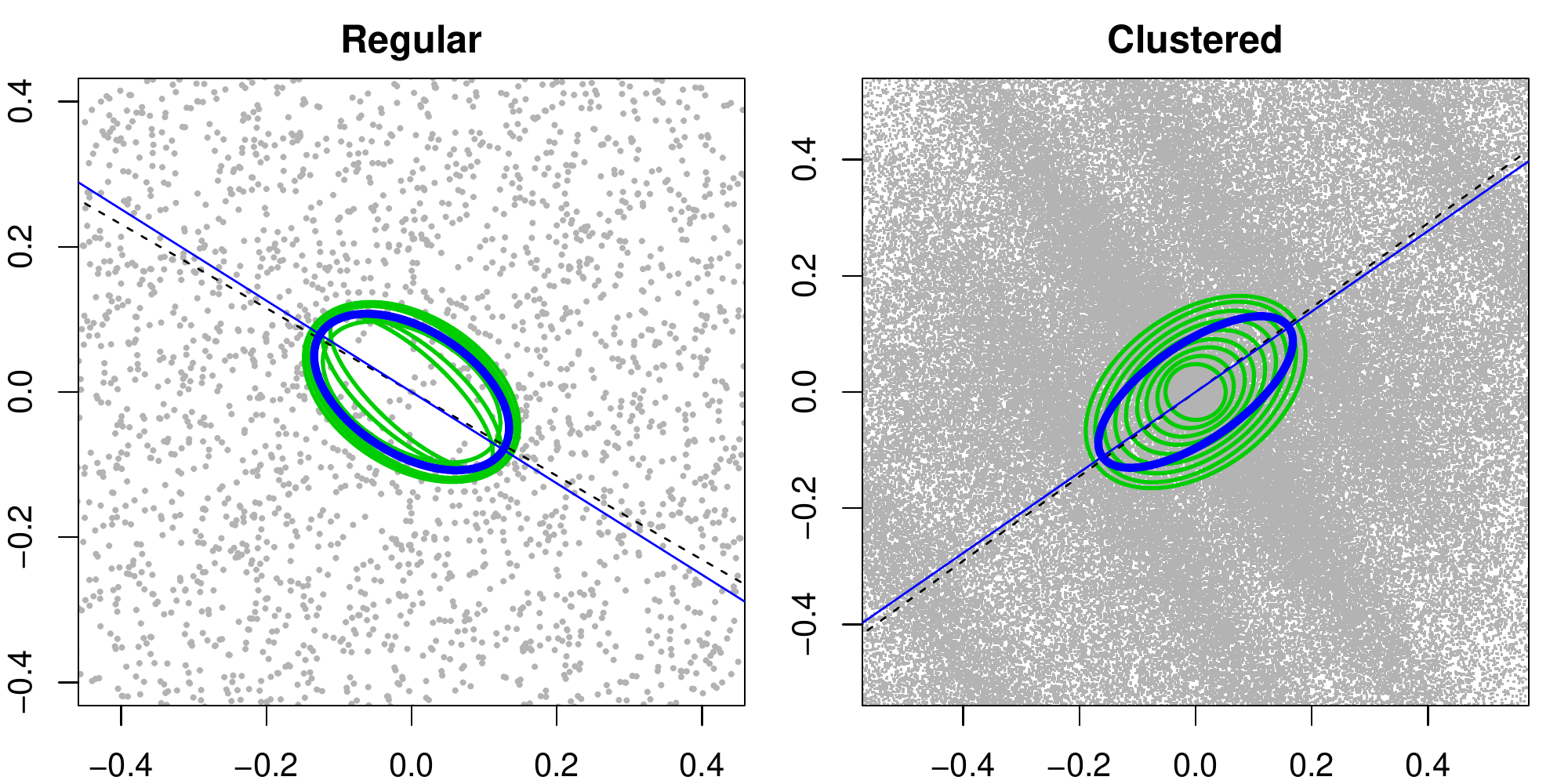}		
	\end{center}
    \caption{Rotation estimation by Fry point ellipsoid fitting for the regular (left) and clustered (right) examples. Figures superimpose over the Fry points the estimated contour ellipses (green), an average ellipse (blue) and its main-axis direction (blue line), and the target direction (dashed line). The target direction for the clustered process is rotated 90 degrees clockwise.}
	\label{fig:ellipsoid}
\end{figure}

Figure \ref{fig:ellipsoid} illustrates the rotation estimation for our two example patterns. For the regular pattern the low-level contours, $l<10$, are most informative as the relevant shape is close to the origin. The first few contours have high variance due to the "noise" pairs at short distances, typical to a Strauss process. For the clustered process the low-level contours $l<10$ are not informative due to the local independence in the process, but at larger contours, $l>100$, the deformation of the cluster shape is captured. Notably, the clustering induces ellipses that have their main-axis perpendicular to the direction of interest.
 

\subsubsection{Inhomogeneous patterns}
\label{sec:SecondOrder_Nonstationary}
If we cannot assume that the intensity function $\lambda(x)$ is constant, the pair correlation needs to be defined as $$g(x,y)= \frac{\rho^{(2)}(x,y)}{\lambda(x) \lambda(y)}.$$ A relaxation of the notion of stationarity is obtained by only assuming that the second-order properties of the point process are translation invatiant, i.e. $g(x,y)=g(x-y)$ does not depend on exactly where $x$ and $y$ are. Then, the summaries can be estimated using the second-order intensity reweighted stationary (SOIRS) process methodology as described by \cite{Baddeley2000} and \cite{VanLieshout2011}. Given $X$ is a SOIRS process, the $\mathcal{K}$-measure is defined as
$$
\lambda^2 \mathcal{K}_{soirs}(B) := \frac{1}{|W|} \E \left(\sum_{x \in \x, y\in X}^{\neq}\frac{1_B(y-x)}{\lambda(x)\lambda(y)} \right).
$$
The corresponding pair correlation function is again defined as the Radon-Nikodym derivative
$$\mathcal{K}_{soirs}(B) = \int_B g_{soirs}(u) du,$$
and the SOIRS extension of the classical K-function would be obtained by setting $B=b(o,r)$. Using e.g.\ a cone or a cylinder, we arrive at the corresponding anisotropic versions of the second order summaries. The estimators are modified by including the varying intensities in the double sums, for example the SOIRS conical pcf estimator becomes
$$\hat{g}_{u,soirs}(r) = \frac{1}{v(\epsilon,r)}\sum_{x,y \in \x}^{\neq} \frac{1(\alpha(y-x,u)<\epsilon)k_h(||y-x||-r)}{|W_x \cap W_y |\hat{\lambda}(x)\hat{\lambda}(y)},$$
see also \cite{Habel2017}.

\section{Spectral analysis}
\label{sec:SpectralAnalysis}
\newcommand{\ftk}{\mathcal{F}}
\newcommand{\pgram}{DFT[\x]}

Fourier transformations and spectral analysis techniques can be used to determine the presence of periodic structures in a spatial point pattern. This yields information on both characteristic scales and directions. The theory for spectral analysis of point patterns was first introduced by \cite{Bartlett1964}. However, at that time the applicability of the method suffered from a lack of efficient tools for computation and visualisation of sample spectra, e.g., Bartlett simply tabulates estimated periodogram values. Consequently, 'virtually no development of spectral methods' was achieved until \cite{Mugglestone1996} published their 'Practical guide to the spectral analysis of spatial point processes'. They argue that the advent of powerful computers and graphics packages has overcome computational limitations such that the method deserves being taken into consideration again. It should be noted that notation and normalization of the Fourier transform in \cite{Bartlett1964} and \cite{Mugglestone1996} differ. The following presentation is closer to \cite{Mugglestone1996} than to \cite{Bartlett1964}.  

We consider a point process $X$ with intensity function $\lambda$ and second-order product density $\rho^{(2)}$. The \emph{covariance density function} is defined as
$$
\gamma(x,y)=\rho^{(2)}(x,y)-\lambda(x)\lambda(y),\quad x, y \in \R^d, x\ne y.
$$
Because $X$ is assumed to be simple with no duplicate points the function above is not well defined at $x=y$, and is set to 0. The \emph{complete covariance density function} is defined as 
$$
\kappa(x,y)=\lambda(x)\delta(y-x)+\gamma(x,y),  
$$
where $\delta(\cdot)$ denotes the $d$-dimensional Dirac delta function. If the process is stationary, the complete covariance density function reduces to 
$$
\kappa(z)=\lambda\delta(z)+\gamma(z),
$$
where $z=y-x$. 

The (Bartlett) \emph{spectral density function} is defined as the Fourier transform of the complete covariance density function, namely 
$$
\ftk(\eta,\omega)=\ftk[\kappa](\eta,\omega)=\int_{\R^d}\int_{\R^d} \kappa(x,y)e^{-i(\eta^T x+\omega^T y)}dx dy, \quad \eta,\omega \in \R^d,
$$ 
where $T$ denotes the transpose and $i=\sqrt{-1}$. The values of $\ftk$ are called the \emph{spectrum} of $X$, and the arguments $\eta,\omega$ are called \emph{frequencies}. In the stationary case, $\ftk(\eta,\omega)$ should depend on $x$ and $y$ only via $z=y-x$. Hence, the equation can be simplified to
\begin{align}
\label{eq:spectraldensity}
\ftk(\omega)&:= \ftk(-\omega, \omega)=\int_{\R^d} \kappa(z)e^{-i\omega^T z}dz
= \lambda + \int_{\R^d} \gamma(z)e^{-i\omega^T z}d z.
\end{align} 


Estimation of $\ftk$ has only been discussed in 2D and when the window $W$ is rectangular with side lengths $l_1$ and $l_2$. In general dimension $d$, let $W$ be rectangular with side lengths $l_j>0, j=1,...,d$. 
The spectrum of the pattern can be estimated by a periodogram \citep{Bartlett1964} which is based on the discrete Fourier transform of the pattern $\x$ 
$$
\pgram(\omega) =  |W|^{-\frac{1}{2}}
\sum_{x\in \x} e^{-i\omega^T x}  = A(\omega) + iB(\omega).
$$
The \emph{periodogram estimator} of the spectrum, and hence the spectral density function, is given by
$$
\widehat{\ftk}(\omega) = \pgram(\omega)\overline{\pgram}(\omega)=A(\omega)^2+B(\omega)^2.
$$
%
This estimator has a bias term which for a rectangular window can be written as
\begin{equation}
\label{eq:bias}
\lambda^2 |W|^{-1}\prod_ {j=1}^d \frac{\sin^2(\frac 1 2 l_j \omega_j)}{(\frac 1 2 l_j \omega_j)^2},
\end{equation}
where $\omega=(\omega_1,...,\omega_d)$. Estimating the spectrum is therefore recommended on frequencies that are integer multiples of $2\pi/l_j$ so that the sine term and hence the bias term become 0. Additional bias comes from taking only a finite integral in \eqref{eq:spectraldensity} (see Section 4 in \cite{Bartlett1964}). The periodogram is asymptotically unbiased for $\omega \neq 0$ when $|W|\rightarrow\infty$ \citep[see also][]{Diggle1987}. To avoid the $\delta$-function spike at $\omega=0$, \cite{Diggle1987} suggest truncation for small $\omega$. They also report that more sophisticated methods for extrapolating the periodogram to $\omega=0$ were tried without success. 

\cite{Bartlett1964} as well as \cite{Mugglestone1996} suggest standardizing the coordinates by replacing each point $x=(x_1, \ldots, x_d)$ by $x'=(n x_1/ l_1, \ldots, n x_d/l_d)$ which is claimed to reduce bias for $\omega \approx 0$. 
However, the benefits of the division by $l_j$ are unclear: for non-cubical windows, it produces an artificial geometric anisotropy which will change the spectra as both the intensity and the product density are affected (see the discussion above as well as \cite{Mugglestone1990}). 
Hence, this step should be used cautiously. 



\begin{figure}[h!]
\centering
\includegraphics[width=0.24\textwidth]{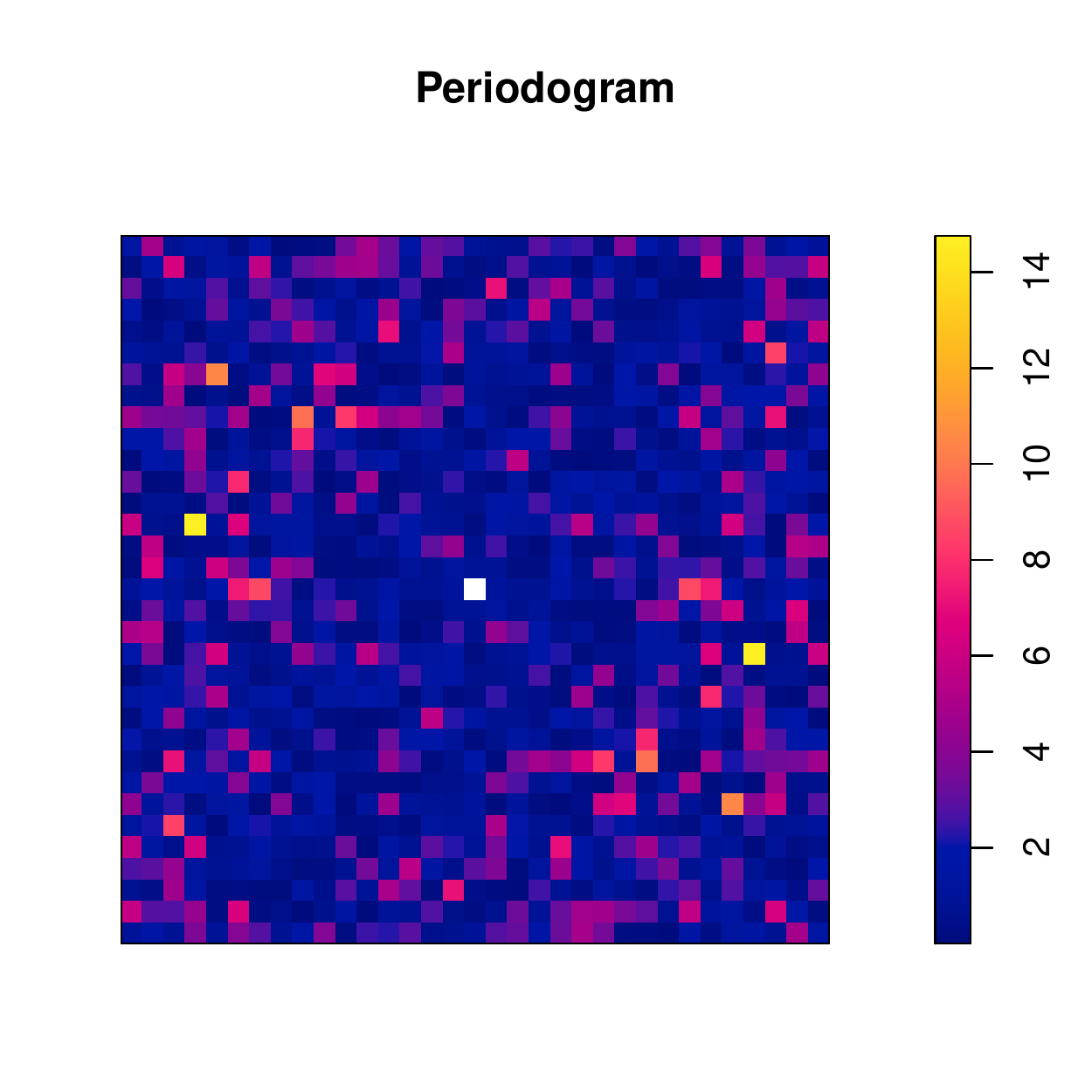} 
\includegraphics[width=0.24\textwidth]{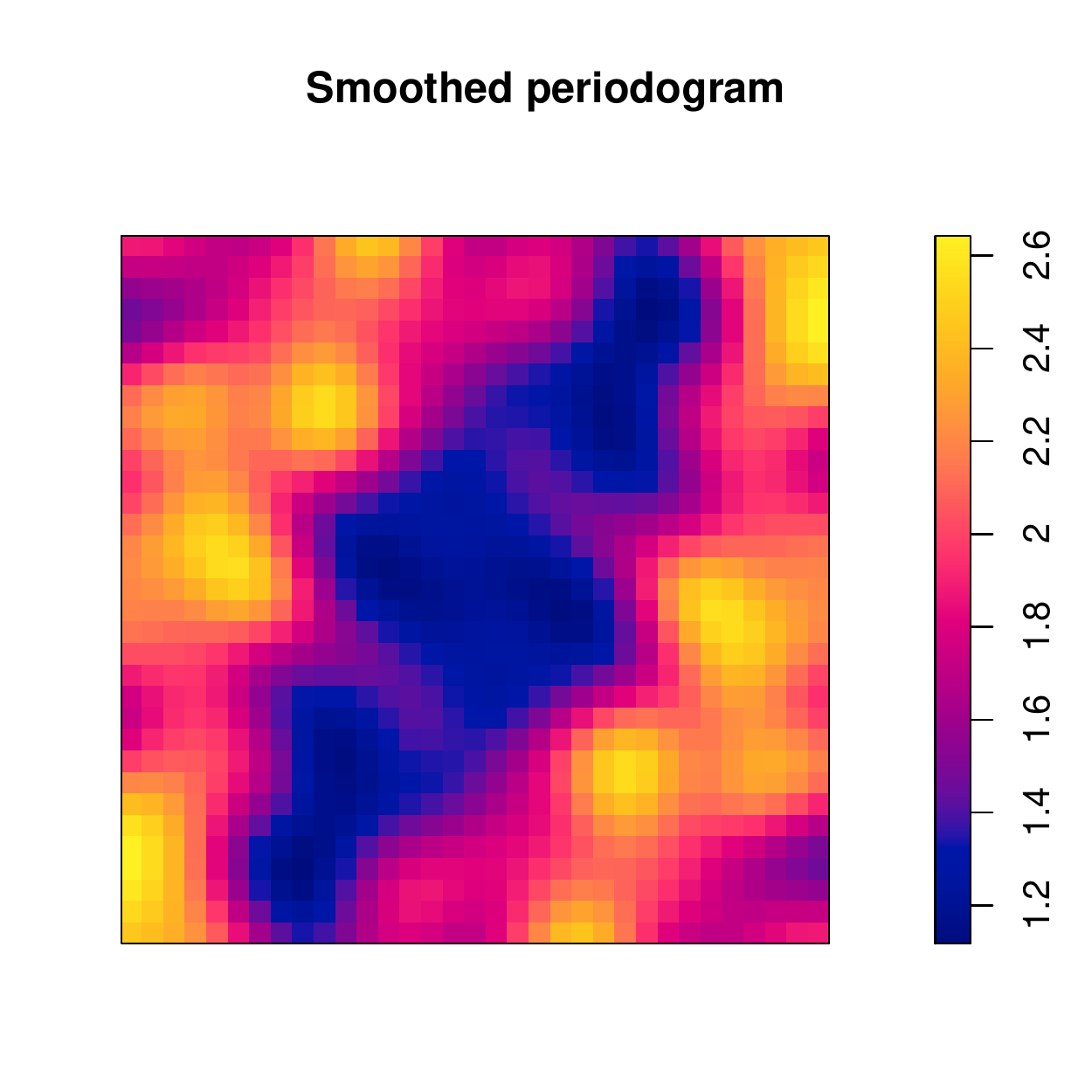}
\includegraphics[width=0.24\textwidth]{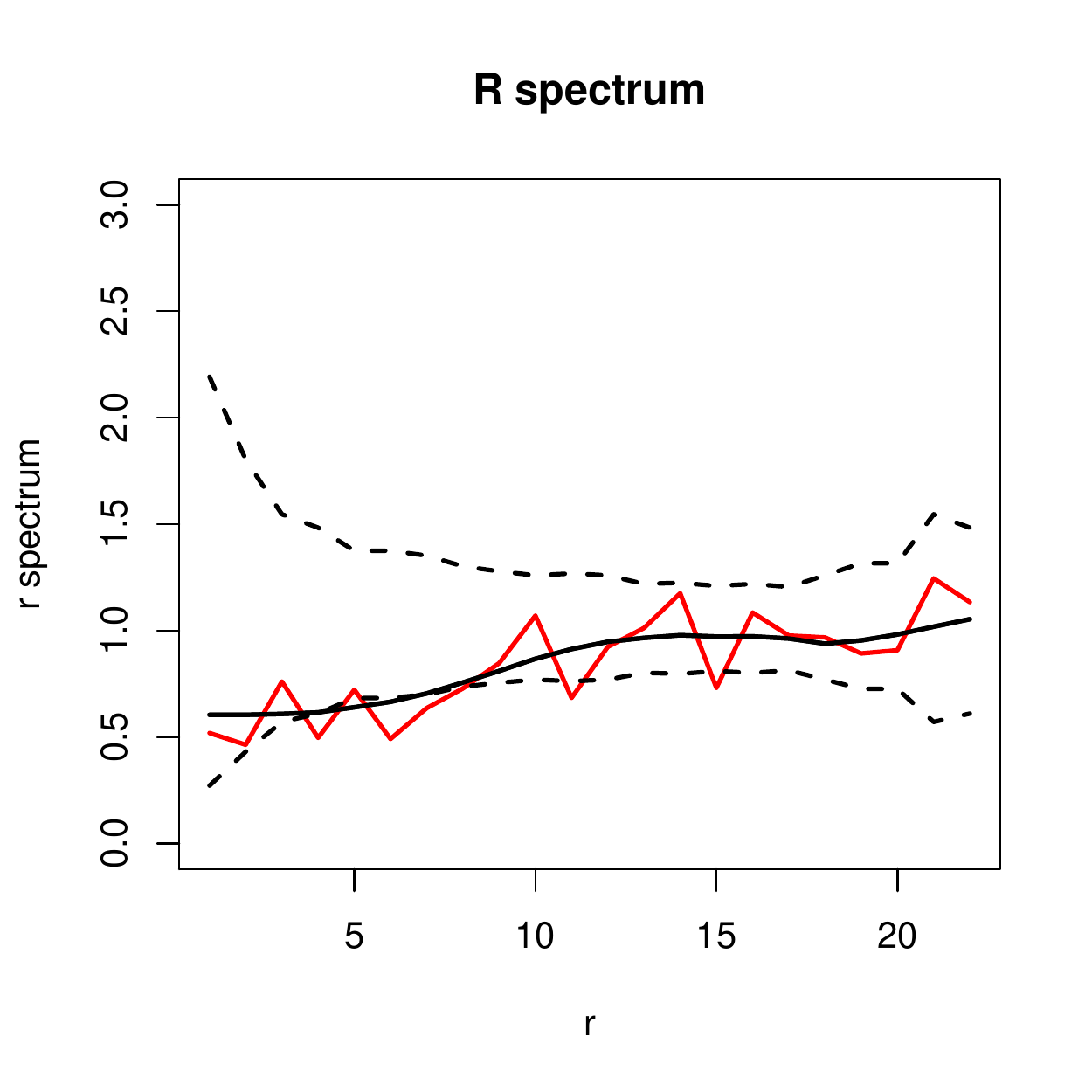}
\includegraphics[width=0.24\textwidth]{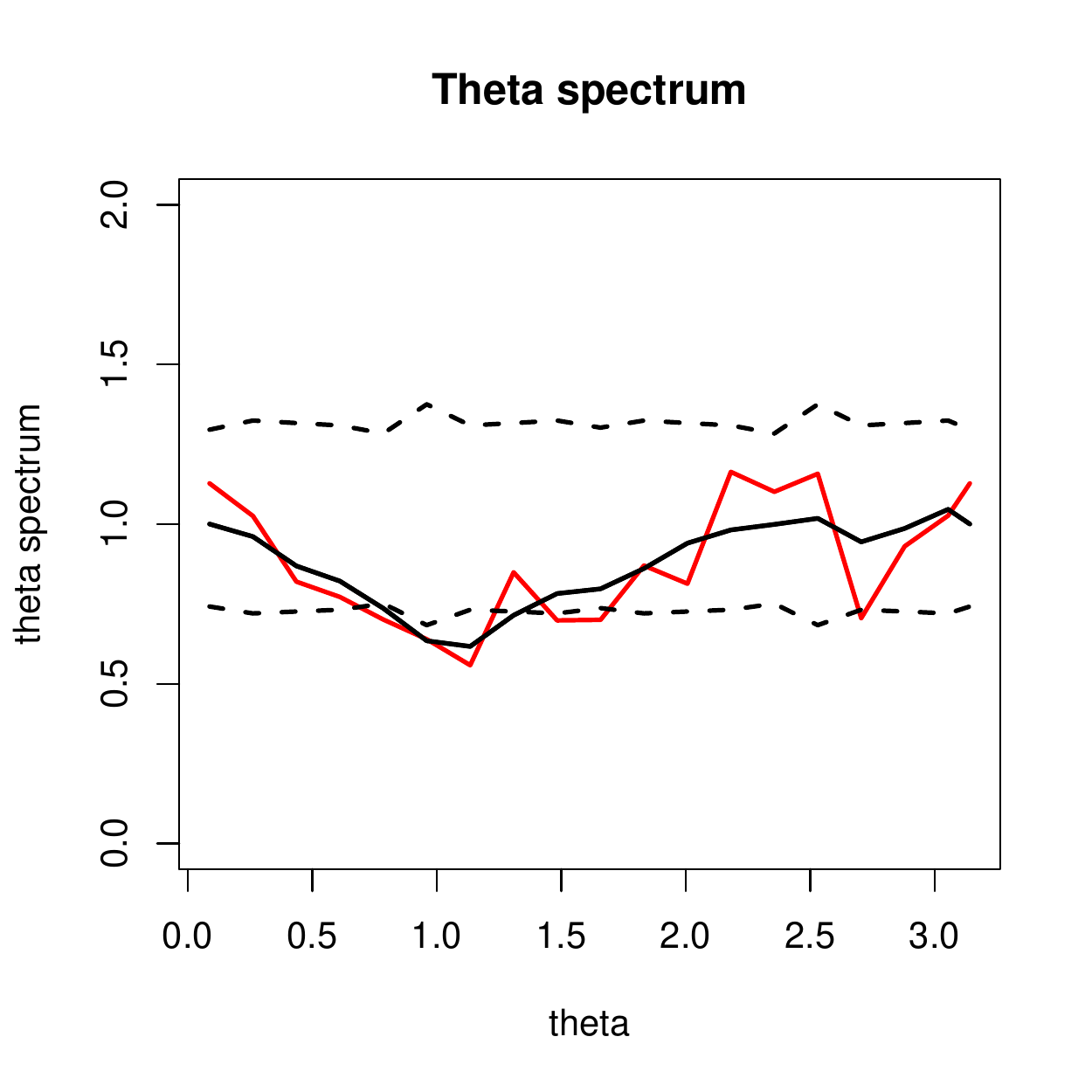}\\
\includegraphics[width=0.24\textwidth]{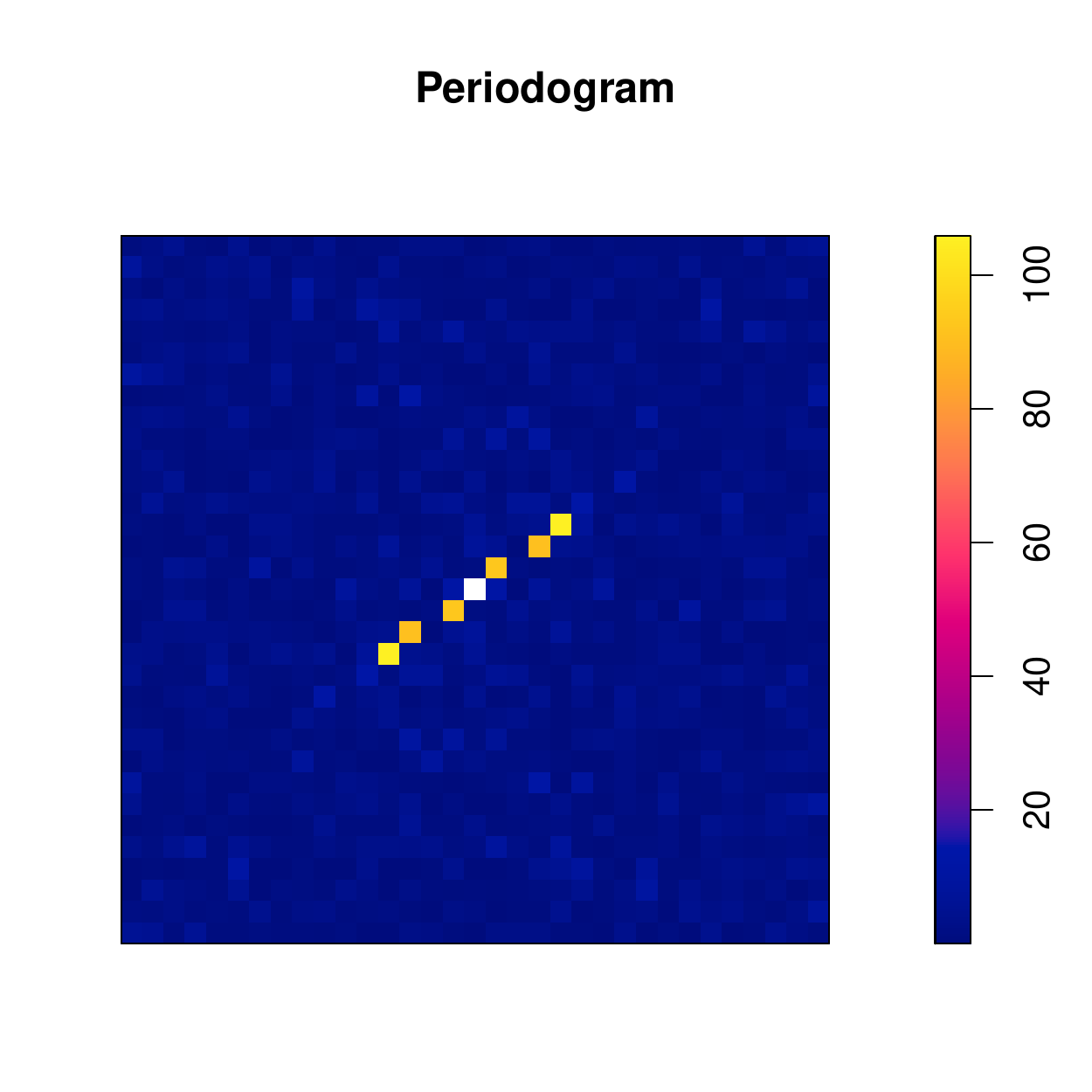}
\includegraphics[width=0.24\textwidth]{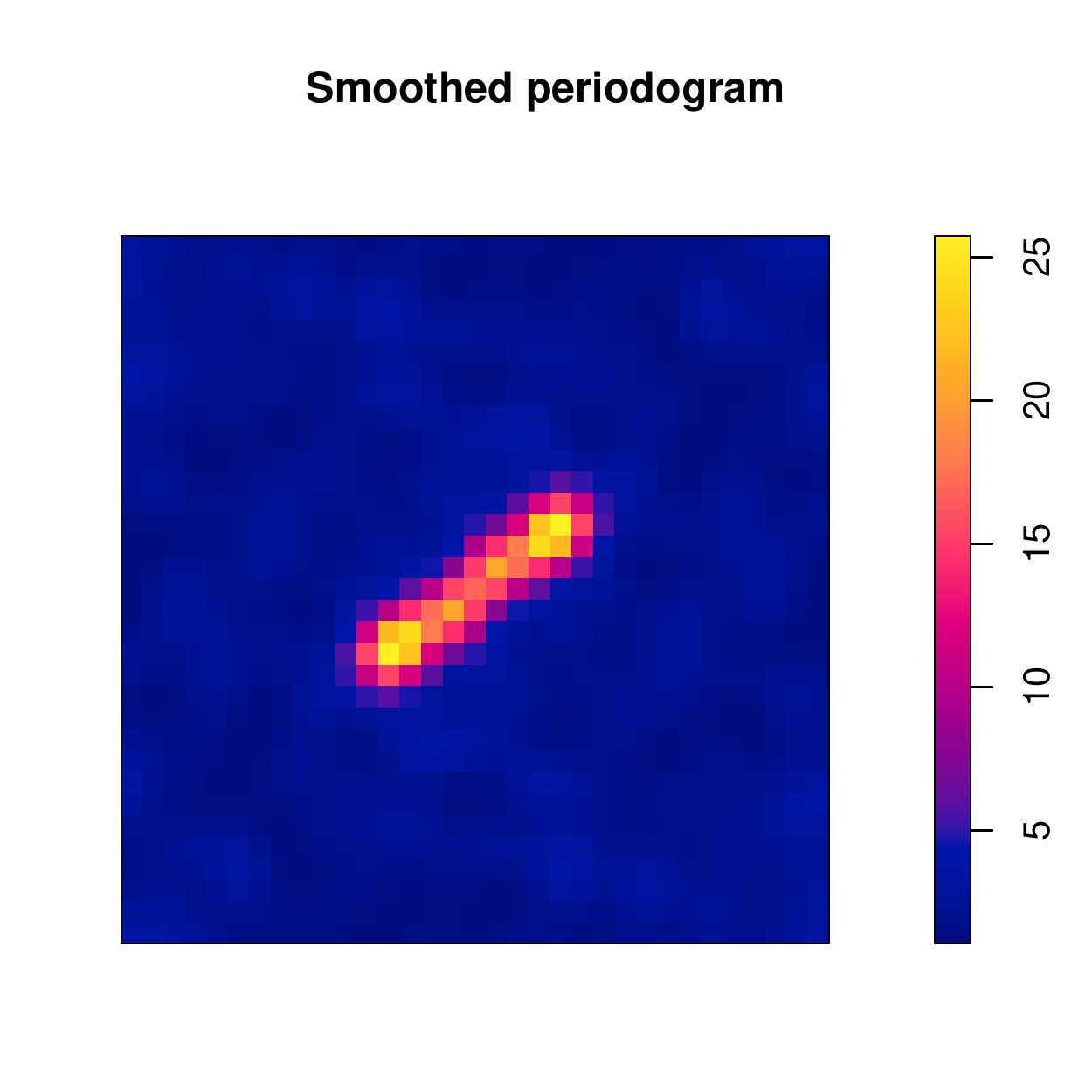}
\includegraphics[width=0.24\textwidth]{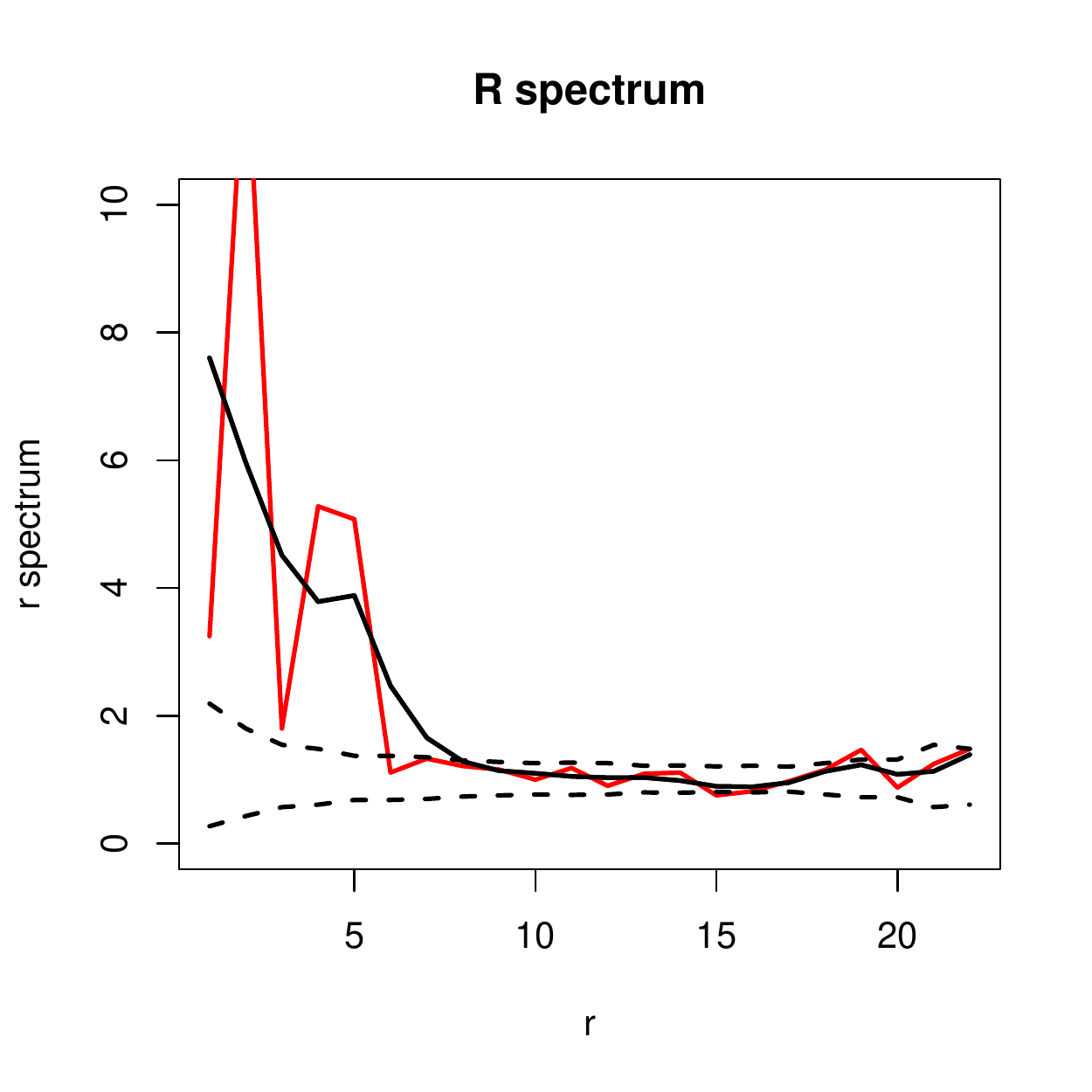}
\includegraphics[width=0.24\textwidth]{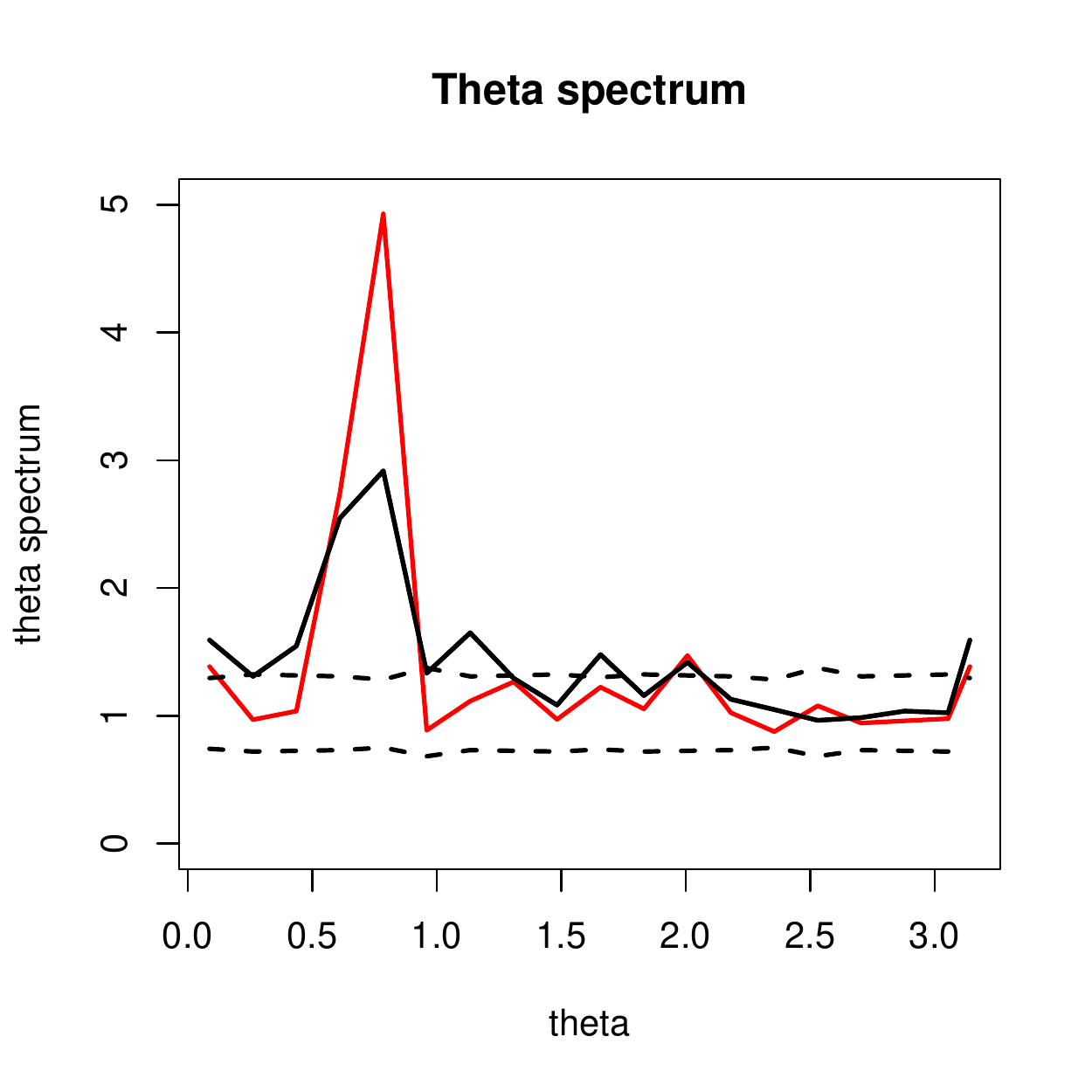}
\caption{ 
Spectral summaries for the regular (top row) and clustered (bottom row) example patterns. From left to right: 
Bartlett's periodogram estimator for the spectrum; Smoothed periodogram; The $R$-spectrum with 95\% confidence intervals (dashed) for the raw (red) and the smoothed (black) periodogram; The $\theta$-spectrum with 95\% confidence intervals (dashed) for the raw (red) and the smoothed (black) periodogram. 
}
\label{fig:spectral-summaries-example}
\end{figure}

The estimates $\widehat{\ftk}(\omega)$ 
are often displayed as a pixel image using the grid of integer factors of the bias-canceling frequencies (Figure \ref{fig:spectral-summaries-example}, left column). The range of suitable integers 
depends on the number of points $n$. As $n$ increases, we expect to obtain more reliable information on the interactions over smaller distances and therefore, higher frequencies. According to \cite{Renshaw1984} and \cite{Mugglestone1990}, in 2D a reasonable range of frequencies is covered by integer factors $\{0,1,...,16\}\times \{-16,-15,...,15\}$, with only the positive values needed for the first (or second) argument as the periodogram estimator is symmetric in the origin. The sizes of the point patterns studied in these papers were roughly between $50$ and $300$ points.

Furthermore, to obtain a consistent estimator of the spectral density the periodogram needs to be smoothed \citep{Bartlett1964}. Concerning time series data, \cite{Chatfield1989} 
discusses the advantages and disadvantages of two types of smoothing techniques, namely smoothing the periodogram using the fast Fourier transform or smoothing the covariance function using lag windows and calculating the periodogram based on the smoothed covariance function. 
He concludes that the smoothed periodogram has better theoretical properties than the one based on the smoothed covariance function. \cite{Kanaan2000}
gives a detailed description of an approach where the periodogram is smoothed by a weighted moving average technique. The same technique is used by \cite{Mugglestone1996b} with the modification that the moving average procedure is repeated several times resulting in a smoother surface, similar to what would be obtained by using a
Gaussian kernel with bandwidth equal to the number of repetitions of the moving average. 

Assuming the biases are not detrimental, the 2D polar coordinate representation of $\widehat{\ftk}(\omega)$ can be useful in anisotropy analysis. It gives two one-dimensional summaries of the periodogram, called the \emph{$R$ spectrum}, $\widehat{\ftk}_R(r)$, and the \emph{$\Theta$ spectrum}, $\widehat{\ftk}_{\Theta}(\theta)$. The $R$ spectrum summarises average periodogram values for ordinates with similar values of $r$ and is used to investigate scales of the pattern under the assumption of isotropy. The $\Theta$ spectrum summarises average periodogram values for ordinates with similar values of $\theta$ and is used to investigate directional features. When evaluating the periodogram on a grid $\{2\pi(p_1/l_1,p_2/l_2)\}$ where $p_1$ and $p_2$ are integers, we consider $r=\sqrt{p_1^2+p_2^2}$ and $\theta=\arctan(p_2/p_1)$ and define the $R$ spectrum as
$$
\widehat{\ftk}_R(r)=\frac 1{n_r}\sum\limits_{r-1<r'\le r}\sum\limits_{\theta}\hat s(\omega_{r'},\omega_{\theta}), \quad r=1,2, \ldots
$$
and the $\Theta$ spectrum as 
$$
\widehat{\ftk}_{\Theta}(\theta)=\frac 1{n_{\theta}}\sum\limits_r\sum\limits_{\theta-5^{\circ}<\theta'\le \theta+5^{\circ}}\hat s(\omega_r,\omega_{\theta'}), \quad\theta=0^{\circ},10^{\circ},...,170^{\circ}.
$$
Here, $\hat s(\omega_r,\omega_{\theta})$ is $\widehat{\ftk}(\omega)$
in polar coordinates, 
and $n_r$ and $n_{\theta}$ are the (chosen) numbers of the periodogram ordinates.
Note that $\widehat{\ftk}(0)$ is not included in the averaging since its asymptotic distribution differs from that of the rest of the periodogram ordinates \citep{Mugglestone1996}.

With the number of points going to infinity, estimates of the spectrum as well as the $R$ and $\Theta$ spectrum are asymptotically distributed as $\chi^2$ random variables. For $\omega \neq 0$, we have: 
$$
\frac{2 \widehat{\ftk}(\omega)}{\ftk(\omega)}\sim \chi^2_{2}, 
$$
and periodogram ordinates for different $\omega$ are asymptotically independent.
Additivity of independent $\chi^2$ random variables implies
$$
\frac 1{n_{r}} \sum\limits_{r'}\sum\limits_{\theta}  \frac{\hat s(\omega_{r'},\omega_{\theta})}{s(\omega_{r'},\omega_{\theta})}\sim \frac{1}{2 n_{r}} \chi^2_{2 n_{r}} 
$$
and
$$
\frac 1{n_{\theta}} \sum\limits_r\sum\limits_{\theta'}  \frac{\hat s(\omega_r,\omega_{\theta'})}{s(\omega_r,\omega_{\theta'})}\sim \frac{1}{2 n_{\theta}} \chi^2_{2 n_{\theta}}. 
$$
 
Under complete spatial randomness (CSR) the periodogram is constant, $\ftk(\omega) = \lambda$ for all $\omega$, which implies 
$$
\widehat{\ftk}_R(r)/\lambda \sim \frac1{2n_{r}}\chi^2_{2n_{r}}
$$
and
$$
\widehat{\ftk}_{\Theta}(\theta)/ \lambda \sim\frac1{2n_{\theta}}\chi^2_{2n_{\theta}}. 
$$
In practice, $\lambda$ is replaced by $\hat \lambda = n/|W|$ which does not change the asymptotic distribution due to consistency of $\hat{\lambda}$ and Slutsky's Lemma. Sometimes it is suggested to generally standardize the periodogram ordinates by dividing by the intensity such that the theoretical value under CSR becomes 1. 
Based on these asymptotics, preferred directions (compared to CSR) in the spectrum can be detected by comparing the scaled estimated $\Theta$ spectrum with appropriate quantiles of the $\chi^2_{2n_{\theta}}$ distribution.

Figure~\ref{fig:spectral-summaries-example} shows the results of a spectral analysis of our two sample patterns.  The raw periodograms and their values after smoothing by an isotropic Gaussian filter are shown together with the resulting $R$ and $\Theta$ spectra. For the linear pattern the standard deviation of the kernel was chosen as $\sigma=1$, so smoothing was done more cautiously than for the regular pattern where we used $\sigma=2$. For the compressed regular pattern, the void ellipse is clearly visible in the periodogram. As compression of a point pattern decreases the period in this direction, the orientation of the ellipse is rotated by $90^{\circ}$ compared to the Fry plot. The compression direction at $\pi/2-\pi/6$ 
is found as the minimum of the $\Theta$ spectrum while the stretch direction is indicated by the maximum at about $\pi-\pi/6$. 
For the clustered pattern, the periodogram shows a bright line perpendicular to the main direction of the clusters. The $\Theta$ spectrum contains a clear peak at the corresponding angle $\pi/5$. 
The direction of the lines is obtained by adding $\pi/2$ to this angle.   

\section{Wavelet analysis} 
\label{sec:Wavelets}


An approach closely related to spectral analysis is obtained by using a wavelet transform instead of a Fourier transform. Wavelet analysis has been used to detect directional properties in 2D spatial point processes first in \cite{Rosenberg2004} and later by D´Ercole, Mateu, and Nicolis in a series of papers \citep{Dercole2010,
MateuNicolis2015,Dercole2013,Dercole2013b,DErcole2014}. All these papers consider anisotropy caused by an increased intensity along directed lines in square windows, the type of anisotropy observed in the Ambrosia dumosa data set (Figure \ref{fig:examplesClustered2D}). 

We first describe the approach of \cite{Rosenberg2004}. Given a realization $\x=\{x_1,...,x_n\}$ of a spatial point process $X$, \cite{Rosenberg2004} starts by selecting one of the points of $\x$ as "specific focal point". The space around the focal point is divided into $360$ angular sectors of width $1^{\circ}$ and directions $\theta_i=i^{\circ}$. Then, the number of further points of $\x$ in each sector is counted and counts from opposite sectors are combined. 


Finally, the count is divided by the area of the intersection of the sector and the observation window yielding the point intensities $\eta(x,\theta_i)$, $\theta_i=0^{\circ} \dots 179^{\circ}$. The 1D discrete wavelet transform of $\{\eta(x,\theta_i)\}$ in direction $\theta$, with some scale parameters $b_k, k=1 \dots m$, and a 1D wavelet function $\psi$ is defined as  
\begin{equation}
\label{eq:RosembergWaveletCoefficients}
W(x, \theta, b_{k})=\frac{1}{b_{k}}\sum_{i=1}^{180}\eta(x, \theta_{i})\psi\left(\frac{\theta_{i}-\theta}{b_{k}}\right). 
\end{equation}

In \cite{Rosenberg2004}, $\psi$ is chosen as the French Top Hat wavelet. According to the paper, the choice of $\psi$ should only weakly affect the results of the analysis. When applying Equation \eqref{eq:RosembergWaveletCoefficients} one should treat $\eta$ as a periodic function in $\theta$ to avoid edge effects in the angle domain. 

The overall variance of the wavelet transform for a given focal point $x$ and direction $\theta$ is defined as
\begin{equation}
\label{eq:RosembergVariance}
P(x,\theta)=\frac{1}{m}\sum_{k=1}^{m}W^{2}(x, \theta, b_{k}),
\end{equation}
which are then averaged over the data points to a directional summary $\bar P(\theta)$. To avoid edge effects, points near the border are excluded from the average.

\begin{figure}[h!]
	\begin{center}
		\includegraphics[width=0.8 \textwidth]{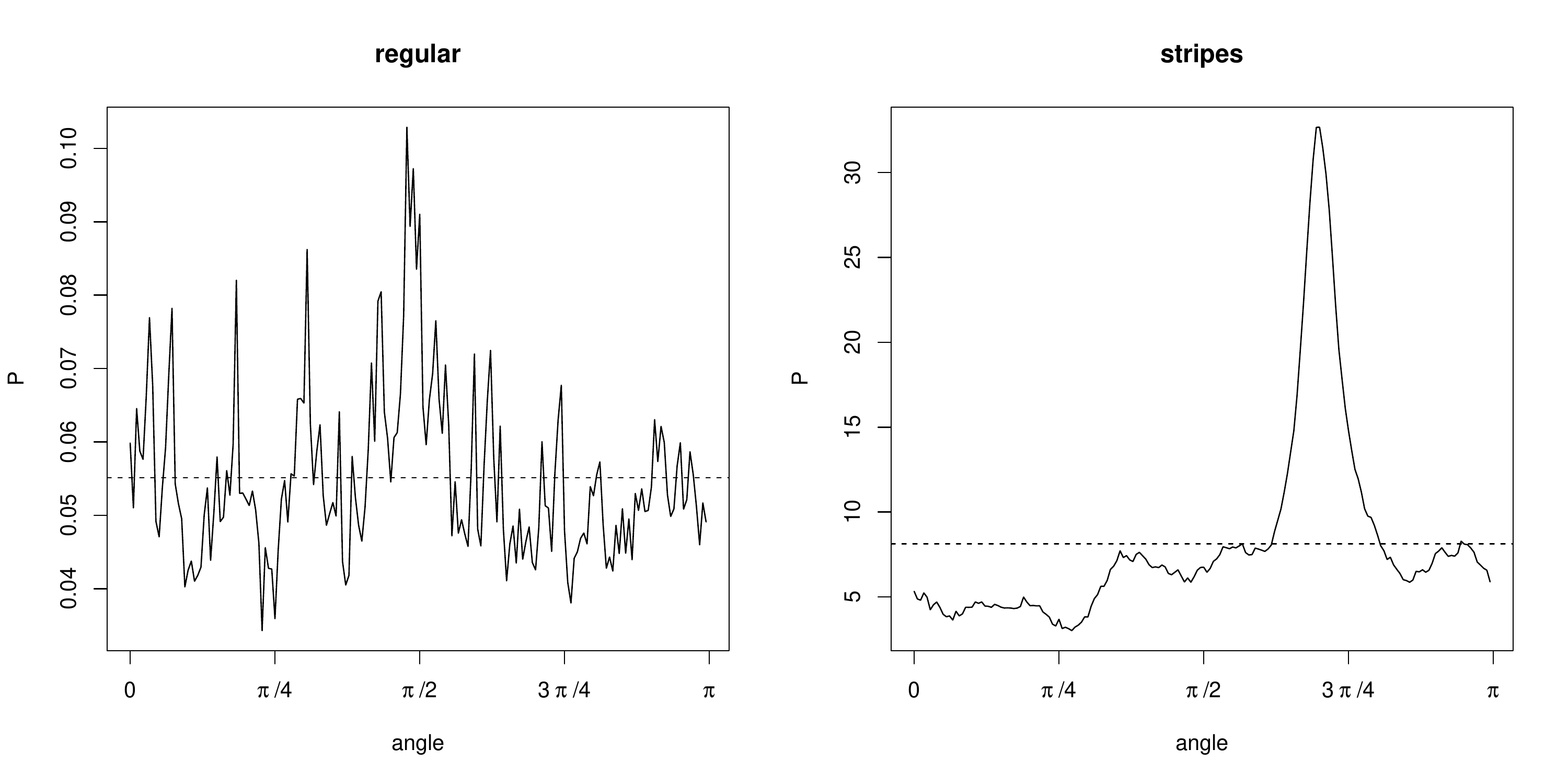}		
	\end{center}
	\caption{The overall variance of Equation \eqref{eq:RosembergVariance} for the regular (left) and the clustered (right) example patterns.}
	\label{fig:Rosemberg}
\end{figure}

The plot of $\bar P(\theta)$ is useful when detecting preferred directions. For example, in the case of directed lines, the peaks of $\bar P(\theta)$ correspond to the directions of the lines. The wavelet analysis as described above is implemented in the software package PASSaGE \citep{MEE3:MEE381}. 
Figure \ref{fig:Rosemberg} shows the average overall variance of Equation \eqref{eq:RosembergVariance} for our two sample patterns with $b_k=1^{\circ} \dots 45^{\circ}$ as used in \cite{Rosenberg2004}. In the clustered case, the plot on the right shows a clear peak in the direction of the clusters around $\frac{\pi}{2}+\frac{\pi}{5}$, while in the regular case (plot on the left), no clear direction is detectable. 

In generalization of Equation \eqref{eq:RosembergWaveletCoefficients}, the 2D directional continuous wavelet transform (CWT) of a square integrable function $f\in L^{2}(\mathbb{R}^{2})$ (with respect to the Lebesgue measure) is defined as
\begin{equation}
F(a,b,\theta)=a^{-1}\int_{\mathbb{R}^{2}}\overline{\psi(a^{-1}R_{-\theta}(x-b))}f(x)dx=:\int_{\mathbb{R}^{2}}\overline{\psi_{a,b,\theta}(x)}f(x)dx,\quad x\in\mathbb{R}^{2},\label{defCWT}
\end{equation}
where $\psi\in{L}^{2}(\mathbb{R}^{2})$ is the mother wavelet of the transform, $a>0$ the scaling parameter, $b\in\mathbb{R}^{2}$ the translation parameter, and $R_{-\theta}$ represents a clockwise rotation by an angle $-\theta$. The wavelet transform is called directional since the mother wavelet $\psi$ is not only translated and rescaled, but also rotated.

Let now $X$ be a (possibly non-stationary) point process with intensity function $\lambda$ observed in a finite window $W$. D´Ercole, Mateu, and Nicolis suggest to apply the directional CWT of equation \eqref{defCWT} to the intensity function restricted to $W$ obtaining the coefficients
\begin{equation}
S(a,b,\theta)=a^{-1}\int_{W}\overline{\psi_{a,b,\theta}(x)}\lambda(x)dx,\quad x\in\mathbb{R}^{2},\label{eq:CWTlambda}
\end{equation}
which can be estimated by estimating $\lambda$ and by discretizing the integral \citep{Dercole2010, MateuNicolis2015}.
An alternative estimator used in \cite{Dercole2013,Dercole2013b, DErcole2014} is given by
\begin{equation}
\hat{S}(a,b,\theta)=a^{-1}\sum_{x\in X \cap W}\overline{\psi_{a,b,\theta}(x)},\label{eq:estimCWTlambda}
\end{equation}
which can be shown to be unbiased by using the Campbell theorem. 

If $X$ is a stationary point process (as considered throughout this paper), the intensity function $\lambda$ is constant. Hence, the transform \eqref{defCWT} will not contain information on possible anisotropies. However, one can consider the squared modulus of $\hat{S}(a,b,\theta)$ given in Equation \eqref{eq:estimCWTlambda} (see \cite{Dercole2013,Dercole2013b,DErcole2014}). In fact, by applying the Campbell formula and its generalisation given in Equation \eqref{eq:Campbell}, it is possible to prove that the expectation of $|\hat{S}(a,b,\theta)|^2$ depends both on the first and second order properties of $X$. As discussed in Section \ref{sec:SecondOrder}, the latter contain directional information also in the stationary case.

Using a similar idea as in Equation \eqref{eq:RosembergVariance}, the square modulus of $\hat{S}(a,b,\theta)$ can be integrated over $b$, obtaining the \emph{scale-angle energy density} \citep{Dercole2013b,MateuNicolis2015} 
\begin{equation}
\nu(a,\theta)=\int|\hat{S}(a,b,\theta)|^{2}db.\label{energy}
\end{equation}
It can be useful in detecting anisotropies since, under isotropy, the energy should be equally distributed in different directions. In \cite{Dercole2013,Dercole2013b} other types of energy densities are described. However, in the stationary case, the scale-angle energy density is the most suitable choice since it exploits the translational invariance of the distribution of $X$.

In Equation \eqref{eq:CWTlambda} the mother wavelet $\psi$ has to be chosen. 
Clearly, to be useful in directional analysis, $\psi$ has to be anisotropic. D´Ercole, Nicolis, and Mateu suggest using the Morlet (or Gabor) mother wavelet, for which the normalized version is given by
$$
\psi(x)=\sqrt{D}(\sqrt{\pi})^{-1}\exp\big(ik_0^Tx\big)\exp\left(-\frac{1}{2}x^TA^TAx\right),
$$ 
where $k_0$ with $||k_0||\ge 5.5$ is the wave vector, $A=\diag(D,1)$ denotes a diagonal matrix, and $D$ is the anisotropy ratio. Note that \cite{Dercole2013,Dercole2013b,DErcole2014} use a slightly different version of the wavelet which is adjusted to integrate to zero. 

\begin{figure}[h!]
	\begin{center}
	    \includegraphics[width=0.6 \paperwidth]{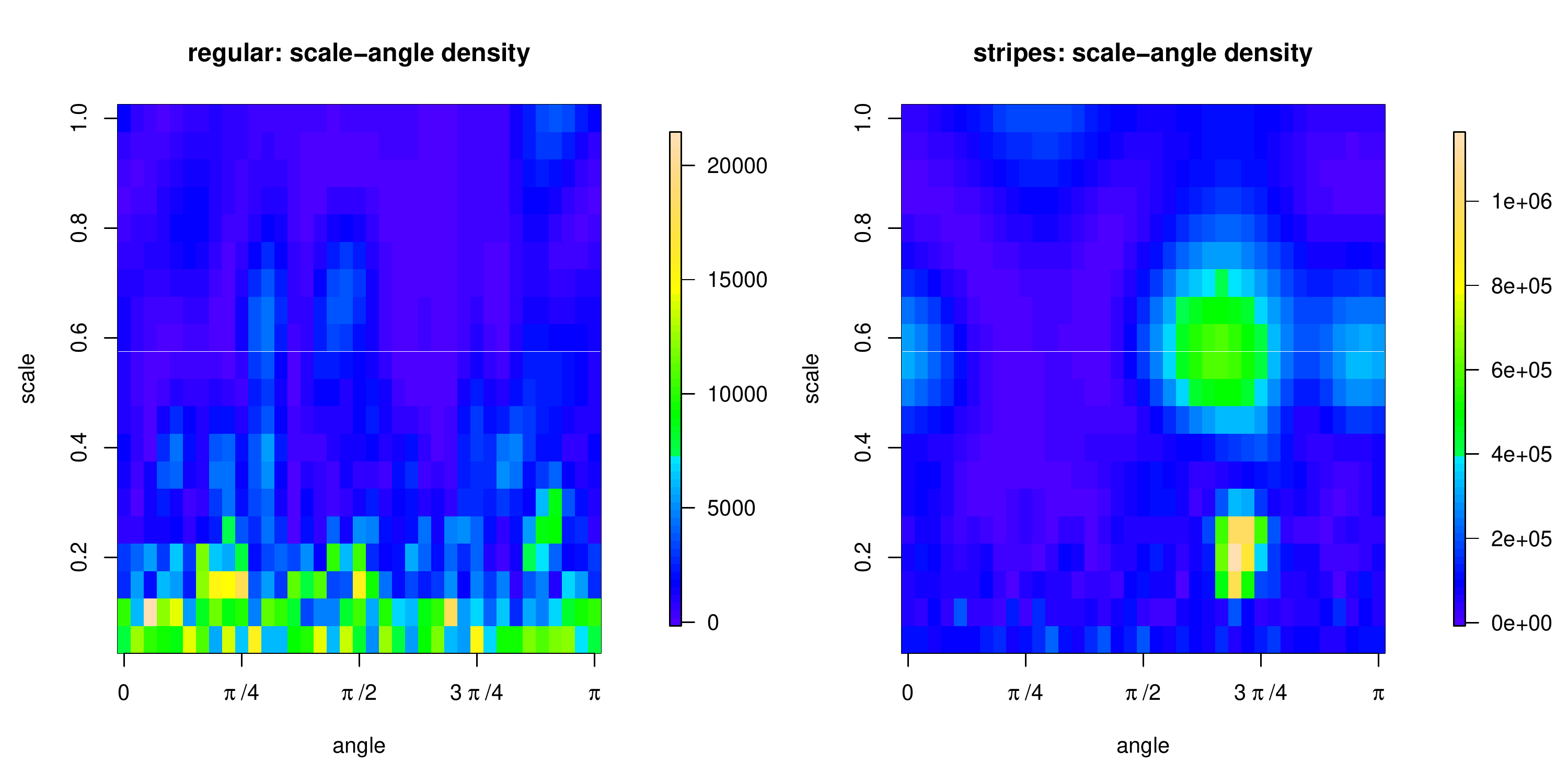}		
	\end{center}
	\caption{Heat-map of the angle-scale energy density map of Equation \eqref{energy} for the regular (left) and clustered (right) example patterns.}
	\label{fig:wavelet}
\end{figure}

Figure~\ref{fig:wavelet} is based on the approach presented by D'Ercole, Mateu and Nicolis and shows the estimated scale-angle energy density. For the analysis, we used the Morlet mother wavelet with $D=0.1$ and $k_0=(0,5.5)$, as suggested in \cite{MateuNicolis2015}. The scale parameter $a$ was chosen between $0$ and the dimension of the window edge length. The energy map for the clustered pattern (right) clearly shows the direction of the clusters around  $\pi/2+\pi/5$ as a bright spot at small scale. However, the energy map for the regular pattern (left) does not give a clear indication of the anisotropy. A different choice of the mother wavelet or adjusting the parameters $D$ and $k_0$ might improve the results.

\section{Testing isotropy}
\label{sec:Testing}
In many practical situations, testing the null hypothesis of isotropy versus anisotropy is of interest. However, the variety of anisotropy mechanisms listed in Section \ref{sec:Mechanisms} shows that there might be no generally best test for this purpose. Most tests for point processes found in the literature consider testing the null hypothesis of complete spatial randomness. Typically, only this case allows for a derivation of theoretical values or distributions of test statistics. As a stationary Poisson process is automatically isotropic, anisotropy in the pattern may result in a rejection of the CSR hypothesis. However, also other deviations from CSR such as regularity or clustering will result in a rejection. Hence, isotropy tests which do not require CSR are of general interest. For many of the statistics considered above it will be impossible to determine the asymptotic distribution of the statistic under a general isotropy hypothesis. In these cases, bootstrapping or the use of replicated data are suitable alternatives to determine critical values of the test statistics.  

\subsection{Nearest neighbour angle uniformity}

A very simple anisotropy test can be based on the observation that under isotropy directions to nearest neighbours should be uniformly distributed on the unit sphere. Hence, a test for this hypothesis could be applied to the nearest neighbour angles. However, as mentioned earlier, the nearest neighbour angles in an observed pattern are not independent. This has to be taken into account when choosing the test. \cite{Konig1992} state that the directional distribution \eqref{eq:dirdistr} (or its estimator \eqref{eq:dirdistrEst}) can be compared to the uniform distribution on the sphere, large deviations indicating anisotropy of the underlying point pattern, but they do not provide any formal test.

As indicated in Section \ref{subsec:nearestneighbour}, nearest neighbour statistics tend to be short sighted. Here, even the distance information is discarded. Hence, tests based on nearest neighbour directions can be expected to have a very small power in general \citep[see also][]{Redenbach2009}.


\subsection{Tests based on summary statistics}

A more promising approach for testing isotropy is to compare directional summary statistics for a set of directions. Such tests are introduced in the papers \cite{Guan2006}, \cite{Wong2016}, \cite{Redenbach2009}, and \cite{Raj16}. 
All approaches are based on similar ideas, but a lot of details are solved in different ways. One of the main questions is how to get hold of the distribution of the test statistics under the null hypothesis.  
The paper \cite{Guan2006} presents a nonparametric approach based on the asymptotic joint normality of the sample second-order intensity function (i.e.\ unnormalised pcf). From that, a test statistic with asymptotic $\chi^2$ distribution is constructed.



Let $\hat{G}=(\hat{\rho}^{(2)}(z_1), \ldots, \hat{\rho}^{(2)}(z_k))^T$ denote a vector of estimates of the second order intensity function $\rho^{(2)}$ at lags $z_1, \ldots, z_k \in \R^d$ with $||z_i|| = ||z_j||$ for all $1 \le i,j \le k $ but representing different directions. As estimator for $\rho^{(2)}$, they use $\hat{\rho}^{(2)}(z_i)= \hat{\lambda^2} \hat{g}(z_i)$, where $\hat{g}$ is a kernel estimator of the pcf as described in Section \ref{subsec:pcf}.

Under the null hypothesis of isotropy, $E[\hat{\rho}^{(2)}(z_i)] =E[\hat{\rho}^{(2)}(z_j)]$ for any pair $z_i, z_j$ of lags. 
The estimated second-order intensities for the different lags are compared via a set of contrasts formed by a full row rank matrix $A$. Then, $AE[\hat{G}]=0$ under $H_0$.  
\cite{Guan2006} consider the test statistic
$$
TS= |W| h^2 (A\hat{G})^T (A \hat{\Sigma}A^T)^{-1} (A \hat{G}),
$$  
where $\hat{\Sigma}$ is an estimate of the asymptotic variance $\Sigma$ of $$|W|^{1/2} h \left(\hat{\rho}^{(2)}(z_1)-E[\hat{\rho}^{(2)}(z_1)], \ldots,\hat{\rho}^{(2)}(z_k)-E[\hat{\rho}^{(2)}(z_k)]\right)^T,$$ which can be obtained using a subsampling approach. 
Under $H_0$ and additional suitable assumptions, the statistic $TS$ is asymptotically $\chi^2_r$ distributed for increasing windows, where $r$ is the rank of $A$.  

The method requires user input on the lags to compare, the bandwidth for $\hat g$, and the subblock size for estimating $\Sigma$. For practical applications, \cite{Guan2006} give the following recommendations:
The subblock size is chosen such that each block contains approximately $cn^{1/4}$ points where $n$ denotes the total sample size. They found $c \approx 0.8$ to be a suitable choice. The simulation study presented in \cite{Guan2006} indicates that the choice of the lags and the bandwidth affects the test size only slightly but has a greater influence on the power. A recommendation by \cite{Guan2006} in the clustered case is to use $||z||$ which is between $1/3$ and $1/2$
of the dependence range (4 times the cluster standard deviation) of $X$, and a recommendation by \cite{Wong2016}
is to use $||z||$ which is between $1/2$ and $2/3$ of the dependence range (2 or 3 times the hard-core distance in the regular case)
while the directions of the lags should be chosen evenly spaced on $[0, \pi]$. The bandwidth $h$ should be chosen such that at least 200 pairs of points are considered in the estimation and disjoint sets of pairs are used to estimate $\rho^{(2)}$ for different directions. They also introduce a data driven method for selecting the bandwidth $h$. However, they admit that this method is computationally intensive and requires a large sample size.

The test presented in \cite{Wong2016} is based on the ratio $F_{r, \psi}(\theta)= K_r(\theta, \psi)/K_r(\pi,0),$ where $K_r(\theta, \psi)$ is the reduced second moment measure of a sector of radius $r$ centred at the origin enclosed between the lines making angles $\psi$ and $\psi+\theta$ w.r.t.\ the $x$-axis. For $\psi=0$, $\theta=\alpha$, $r_1=0$, and $r_2=r$ this corresponds to the second order orientation distribution $F_K(\alpha)$ discussed in Section \ref{subsec:nearestneighbour}. Under the null hypothesis of isotropy, $F_{r, \psi}(\theta)$ is the distribution function of the uniform distribution on $[0, \pi]$. The test is based on the Kolmogorow-Smirnow statistic
$$
d_{r, \psi}= \sup_{\theta \in [0, \pi)} \left| \frac{\theta}{\pi} -\hat{F}_{r, \psi}(\theta) \right|,
$$
whose maximum w.r.t.\ $\psi$ is considered, i.e. 
$$
T_r= \sup_{\psi \in[0, \pi)} d_{r, \psi}.
$$

As a suitable choice of $r$, \cite{Wong2016} recommend a half to two thirds of the dependence range of the point process, see the comment above. In contrast to the approach by \cite{Guan2006}, the asymptotic distribution for the test statistic $T_r$ under the null hypothesis of isotropy is not known. Hence, a critical value of the test statistic is determined by 
using the reconstruction algorithm of \cite{Tscheschel2006} 
to simulate 
isotropic versions of the data such that some predefined 
isotropic summary statistics are close to the ones of the real data. 
In a simulation study, the approach is applied to simulated realisations of regular and clustered anisotropic point patterns and compared to the asymptotic approach by \cite{Guan2006}. An advantage of the bootstrap test is that there is no need to choose a bandwidth. Additionally, according to the authors, the power of the bootstrap test is more robust to the choice of the user-specified parameters than the power of the test by \cite{Guan2006}.

While the two tests discussed so far have been introduced for the 2D case, testing anisotropy in 3D point patterns has been considered in \cite{Redenbach2009}. The test statistics used there are based on the conical K-function and the directional nearest neighbour distance distribution (see Section \ref{sec:SummaryStatistics}). Again, the test is nonparametric in the sense that no parametric model is assumed for the data. To derive critical values for the test, replicated data are used. 
The particular type of anisotropy investigated in the paper is geometric anisotropy generated by a transformation of an isotropic pattern with a matrix $T= \diag(1/\sqrt{c}, 1/\sqrt{c}, c)$. Hence, the difference of the summary statistics for the $x$- and $y$-direction can be used to describe the behaviour under the null hypothesis. Test statistics used are
$$
T_{xy,i} = \int_{r_1}^{r_2} |\hat{S}_{x,i}(r) -\hat{S}_{y,i}(r)| dr, i=1, \ldots, n 
$$
and
$$
T_{z,i} = \min \left(\int_{r_1}^{r_2} |\hat{S}_{x,i}(r) -\hat{S}_{z,i}(r)| dr , \int_{r_1}^{r_2} |\hat{S}_{y,i}(r) -\hat{S}_{z,i}(r)| dr \right), i=1, \ldots, n. 
$$
Here, for instance, $\hat{S}_{x}$ is an estimate of a directional summary statistic pointing to the $x$-axis. When using these test statistics there is no need for choosing a bandwidth. Instead, the opening angle of the cone or the cross-section half-length of the cylinder needs to be chosen. Furthermore, the method is sensitive to the choice of the integration limits which is equivalent to the choice of the lags in the two previous methods. For hardcore processes, \cite{Redenbach2009} recommend choosing the right limit $r_2$ slightly larger than the hardcore distance. They perform a simulation study comparing the powers of tests based on the conical K-function as well as the local and global directional nearest neighbour distance distributions. When using the optimal integration range, the tests based on the conical K-function had the highest powers. In contrast, the test based on the local nearest neighbour distance distributions was more robust to the choice of the integration range than the test based on the K-function. The test based on the global nearest neighbour distance distributions behaved poorly.

Ellipsoids fitted to the Fry plot can also be used to construct tests against isotropy as discussed in \cite{Raj16}. The least squares approach provides asymptotically normal estimators of the coefficients in the quadratic model space, by which one can simulate and compute Monte Carlo confidence intervals for chosen functions of the parameters. Especially, the equality of the semi-axes of the ellipsoid can be assessed. The contrast of interest in 2D is $"a_1 - a_2 = 0"$, where $a_i$ is the $i$th semi-axis length. In 3D, one can check either the contrasts $"a_1-\frac{1}{2}(a_2+a_3)=0"$, $"a_2-\frac{1}{2}(a_1+a_3)=0"$ or $"a_3-\frac{1}{2}(a_1+a_2)=0"$. 

As reported in \cite{Raj16}, in the case of a regular process the contour ellipses $E_l$ are converging towards a circle/sphere with increasing $l$ and therefore, $E_l$ overestimates the "roundness" of $T$ as $l$ increases. The test is expected to be conservative, i.e.\ the test based on $E_l$'s with large $l$ is not expected to be very powerful. In the clustered case, the ellipsoids tend to be circular for small $l$. Therefore, the test should be based on a reasonably small $l$ in the regular case and reasonably large $l$ in the clustered case. How large $l$ should be depends on the process and its intensity. The recommendation in \cite{Raj16} is to choose three or four different values for $l$ and to check the confidence intervals for these values. Note that the results need to be adjusted due to the multiple testing.

Performance of the isotropy test was investigated in a simulation study both in 2D and in 3D. First, realizations of a stationary and isotropic Strauss process were simulated and then, the realizations were compressed to obtain anisotropic patterns. Several regularity parameters and compression strengths were used. As expected, the power of the test increased with regularity and compression both in 2D and in 3D. The test has a very small power when the compression factor is larger than 0.9, i.e.\ the pattern is very close to being isotropic.

\subsection{Spectral tests}
\cite{Mugglestone2001} propose five tests against CSR based on the periodogram. None of them is, however, based on the $\Theta$ spectrum that could show deviations from isotropy. According to \cite{Mugglestone1996}, for general testing of isotropy without assuming CSR, one could scale the $\Theta$ spectrum by the average periodogram value rather than $\lambda$ to investigate whether spectral power is distributed evenly across the frequency angles. Details are not given.



\subsection{Tests based on wavelet analysis}
An anisotropy test based on wavelet transforms is proposed in \cite{MateuNicolis2015}. They estimate the wavelet coefficients $S(a,b,\theta)$ of Equation \eqref{defCWT} (Section \ref{sec:Wavelets}) for a given selection of directions $\theta_i  \in (0, \pi]$, $i=1, \ldots, m$, scales $a_j$, $j=1, \ldots, L$, and locations $b_k$, $k=1, \ldots, N$, in the domain of $\lambda$. They then compute a discretized version of the scale-angle energy density $\nu(a,\theta)$ in Equation \eqref{energy}, summing up the modules of the wavelet coefficients over the positions $b_k$. Finally, they consider the following sums:
$$
T(\theta_i)= \frac 1 L\sum_{j=1}^L \log(\nu(\theta_i, a_j )), \quad i=1, \ldots, m.
$$
Under isotropy, this test statistic should not depend on the direction. However, the exact distribution of $T(\theta_i)$ is unknown. In practice, \cite{MateuNicolis2015} approximate the distribution by the empirical distribution of the $T(\theta_i)$ for an isotropic parametric model fitted to the data. For the directions, they choose $\theta_i = i^{\circ}$, $i=1, \ldots, 180.$ The grids for $a_j$ and $b_k$ are not explicitly specified. 
In a simulation study, to test the null hypothesis of isotropy they introduce separate tests for each direction.
Anisotropy tests based on energy densities different from the scale-angle energy density are described in \cite{DErcole2014}.

\section{Discussion}

In this paper, we have given an overview of methods for directional analysis of unmarked stationary point processes covering methods based on Fry points and summary statistics, spectral analysis, and wavelet analysis. We have illustrated how the methods can be used to detect isotropy, to estimate favorable directions in the data and to test for isotropy. The methods based on the nearest neighbor and second-order summary statistics are described in 2D and 3D and the methods based on wavelet analysis in 2D. Spectral analysis, except the polar coordinate representation, is presented in general dimension even though in the literature the presentation has been limited to 2D. We have considered  two simple simulated examples, a compressed and rotated realization of a Strauss process to represent geometric anisotropy, and a pattern with clustering along parallel stripes to represent 1st order anisotropy, to illustrate how the results of the analyses based on the different methods would typically look like. Based on the plots in Sections \ref{sec:SummaryStatistics}, \ref{sec:SpectralAnalysis}, and  \ref{sec:Wavelets}, we can make some observations concerning the suitability of the methods in these two archetypal cases. However, a much more thorough study than the one presented here would be needed to draw any general conclusions.



Our review does not cover all aspects of anisotropy that have been discussed in the literature. First, we have concentrated on point patterns that are realizations of stationary point processes. In the literature, directional analysis based on wavelets has been focused on situations where anisotropy can be detected from the intensity. Typically, the patterns that have been analyzed have a higher intensity of points along a line or along two perpendicular lines, and stationarity has not been assumed. Second, we have not given a thorough overview of anisotropic stationary point process models even though some of them are mentioned in the paper.  Finally, there is some literature on anisotropy in marked point processes, namely orientation of marks and anisotropic distribution of points (see e.g.\ \citealt{PenttinenStoyan1989, Stoyan1991, StoyanBenes1991}) which is not covered here.


We would like to finish by pointing out some directions for future research. Extension of the methods to higher dimensions, for example finding a polar coordinate representation in spectral analysis and a proper treatment of rotations in the wavelet approach, would be desirable. More in-depth analysis of the statistics of the summary functions would be beneficial as directional sub-sampling effectively reduces sample sizes, making user's input on estimation  parameters more influential.  Related to this, testing for isotropy could be made more objective as the current tests depend heavily on the user providing good input parameters, for example the bandwidth and lag vectors in the asymptotic test by \cite{Guan2006}, distance $r$ in the test based on stochastic reconstruction by \cite{Wong2016}, opening angle and integration limits in the test based on replicates by \cite{Redenbach2009}, and opening angle and contour level in the ellipsoid based test by \cite{Raj16}. The relevant mathematics might turn out to be overtly challenging, in which case guidelines could be improved by diverse simulation experiments. The current tests based on spectral and wavelet analyses require some specified isotropic null model, which can be difficult to specify especially in 3D. Therefore, general tests that do not require any pre-specified model for the location process 
would be welcome. An interesting question related to the modelling of anisotropic point patterns is whether the linear transformation is the only useful, generally applicable anisotropy mechanism if the location process and anisotropy are considered separable. 

\section{Acknowledgement}
This work was supported by the Deutsche Forschungsgemeinschaft (DFG) in the framework of the priority programme "Antarctic Research with comparative investigations in Arctic ice areas" by a grant RE 3002/3-1, by the Knut and Alice Wallenberg Foundation (KAW 2012.0067), and by Swedish Foundation for Strategic Research (SSF AM13-0066). We would also like to thank the two anonymous referees for very valuable
comments.

\bibliographystyle{plainnat}
\bibliography{directed_summaries_review}

\end{document}